# Column-density estimation from the equivalent widths of absorption doublets

Avshalom Badash ★ and Doron Kushnir

*Department of Particle Physics and Astrophysics, Weizmann Institute of Science, Rehovot 76100, Israel*

Accepted 2026 October 6. Received 2026 September 12; in original form 2026 August 1

## ABSTRACT

Na ID absorption is widely used to estimate the dust extinction towards Galactic and extragalactic sources, but the inferred column densities are unreliable once the lines saturate. We present an analytical framework for estimating column densities from the equivalent widths (EWs) of absorption-line doublets. The method requires only that the two doublet members be measured separately – the velocity structure need not be resolved – so a resolving power $\lambda/\Delta\lambda \gtrsim 10^3$ suffices for Na ID, two orders of magnitude below velocity-resolved methods. For the Gaussian curve of growth, the classical inversion reduces to a universal saturation correction depending only on the doublet ratio, $R = \mathrm{EW}_2/\mathrm{EW}_1$: At fixed $R$, the inferred column density is linear in the measured EW. The formulation yields the exact Gaussian inversion, $N_{\tilde{\kappa}}$, which provides the Doppler parameter; a closed-form second-order approximation; and a strict lower bound, $N_{\rm min}$, valid for an arbitrary velocity structure. We recommend reporting $N_{\rm min}$: an optimal, assumption-free lower bound, within a few per cent of $N_{\tilde{\kappa}}$ for $R \gtrsim 1.4$ and at most a factor 3.5 below it at $R = 1.1$. The ratio $N_{\tilde{\kappa}}/N_{\rm 2nd}$ provides a per-object saturation diagnostic. We validate the framework against published profile-fitting Na ID column densities towards Galactic sightlines and Type Ia supernovae, using only the integrated EWs. An explicit closed-form formula converts the two EWs into an extinction estimate through the observed $\log N(\mathrm{Na\,I})$–$A_V$ relation. A public implementation with full uncertainty propagation accompanies the paper.

## 1 INTRODUCTION

Interstellar absorption lines provide one of the most direct probes of the column densities, kinematics, and physical conditions of gas along astrophysical lines of sight. Two optical resonance doublets have played a particularly central role in studies of the interstellar medium: Na ID at $\lambda\lambda \simeq 5890, 5896$ Å and Ca II H&K at $\lambda\lambda \simeq 3934, 3968$ Å (e.g. C. E. Albert et al. 1993; K. R. Sembach, A. C. Danks & B. D. Savage 1993; D. E. Welty, L. M. Hobbs & V. P. Kulkarni 1994; J. V. Smoker et al. 2003; I. Hunter et al. 2006; R. F. Carswell & J. K. Webb 2014). These lines are strong, accessible from the ground, and have oscillator strengths, $f$, whose ratios are close to two, making them particularly useful diagnostics of saturation ($f_{\rm Na\,ID} \simeq 0.641,\ 0.320$ and $f^{\rm H\&K}_{\rm Ca\,II} \simeq 0.627,\ 0.312$; D. C. Morton 2003). Crucially, the two members of each doublet share the same lower (ground) level, so they probe the same absorbing population and differ effectively only in their oscillator strengths. In the optically thin limit, the equivalent width (EW) of a transition is proportional to the column density. Once the line becomes saturated, however, the EW grows more slowly with column density, and an estimate based on the thin-line approximation can severely underestimate the true column. This is the classical curve-of-growth (COG) problem. In particular, on the flat (saturated) part of the COG the same measured EW can correspond to very different column densities, depending on the velocity distribution of the absorbing gas – which is unknown unless the line profile is fully resolved.

The use of resonance-line doublets to infer interstellar column densities dates back to the pioneering work of O. C. Wilson & P. W. Merrill (1937), who showed that the observed Na ID doublet cannot be explained by a homogeneous Maxwellian absorber or Galactic rotation alone, and argued instead that its saturation behaviour arises from discrete interstellar clouds with random velocities, interpreted as the gaseous counterparts of the Galactic dust clouds. Building on this cloud picture, B. Strömgren (1948) formalized the Gaussian COG analysis basis by tabulating the relation between optical depth, EW, and doublet ratio defined as

$$R = \frac{\mathrm{EW_{strong}}}{\mathrm{EW_{weak}}}, \tag{1}$$

where strong and weak denote the larger and smaller oscillator strengths, respectively. Strömgren demonstrated that a line of sight intersecting multiple unresolved Gaussian clouds with identical intrinsic Doppler widths can be treated using an effective Doppler parameter.

The Gaussian COG framework was subsequently adopted in studies of interstellar Na I and Ca II H&K absorption (e.g. L. M.

★ E-mail: avshalom.badash@weizmann.ac.il

Hobbs 1969) and became the standard approach described in textbooks (D. Mihalas 1978; L. Spitzer 1978). Throughout this classical development, however, the inversion remained numerical: The observed doublet ratio was first converted into the corresponding Gaussian COG through tabulated relations, from which the line-centre optical depth, Doppler parameter, and column density were recovered. P. Nachman & L. M. Hobbs (1973) generalized the doublet-ratio method to multicloud sightlines and showed that unrecognized velocity structure can bias the inferred column density by up to a factor of $\sim$10. Motivated by the growing complexity of the interstellar velocity structure, E. B. Jenkins (1986) revisited the validity of the classical Gaussian COG for unresolved absorbers. Using analytical expansions and Monte Carlo simulations, he showed that the composite EWs of large ensembles of Gaussian absorption components spanning broad distributions of optical depths and Doppler parameters closely reproduce the classical single-component Gaussian COG, strengthening the conclusion that standard doublet-ratio analysis can recover the total column density with surprisingly small systematic errors over a broad range of physically relevant conditions.

Modern analyses of interstellar absorption increasingly rely on methods that exploit the full line profile rather than integrated EWs. Since the work of L. Spitzer & W. A. Morton (1976), profile-fitting techniques have modelled absorption features as a superposition of individual velocity components, each characterized by a column density, velocity centroid, and line-broadening parameters, convolved with the instrumental line-spread function. These methods are implemented in widely used software packages such as VP FIT (R. F. Carswell & J. K. Webb 2014); J. K. Webb, R. F. Carswell & C.-C. Lee 2021), VOIGTFIT (J.-K. Krogager 2018), and ASTROCOOK (G. Cupani et al. 2020), and have been applied extensively to studies of interstellar and circumgalactic absorption in both Galactic and extragalactic environments (e.g. K. R. Sembach et al. 1993; I. Hunter et al. 2006; M. M. Phillips et al. 2013, hereafter P13). An alternative approach is the apparent optical depth (AOD) method introduced by B. D. Savage & K. R. Sembach (1991), which derives the column density directly from the observed absorption profile without assuming an explicit velocity-component model. Each velocity bin of the observed profile is converted into an apparent optical depth, $\tau_a(v) = \ln[I_c(v)/I(v)]$ (with $I_c$ the continuum), and thence into an apparent column density per unit velocity, $N_a(v) \propto \tau_a(v)/(f\lambda)$, whose integral over the profile yields the total column density without any assumption about the number or shapes of the velocity components. An unresolved saturated structure leads to an underestimate of the true column density, but it can be diagnosed empirically by comparing the apparent column-density profiles of transitions with different oscillator strengths. B. D. Savage & K. R. Sembach (1991) further presented an empirical correction that relates the discrepancy between the integrated apparent column densities of the two transitions to the expected underestimate of the weaker transition. E. B. Jenkins (1996) subsequently developed a local velocity-dependent correction that applies the saturation correction directly to the apparent optical-depth profile. The AOD method has since been applied to interstellar absorption-line studies (e.g. K. R. Sembach et al. 1993; A. J. Fox, B. D. Savage & B. P. Wakker 2005).

Both the profile-fitting and AOD approaches require spectra that resolve the absorbing gas's velocity structure. Cold interstellar clouds have Doppler parameters of $b \sim 1$–$10\,\mathrm{km\,s^{-1}}$ (e.g. D. E. Welty et al. 1994), so resolving their velocity structure requires a high spectral resolving power of $\lambda/\Delta\lambda \sim c/b \approx 3\times10^4$ to $3\times10^5$. In contrast, we consider the complementary problem of inferring the column density from the integrated EWs of a resolved doublet. The only observational requirement is that the two members of the doublet be separated from each other; for the Na ID doublet, whose members are $\simeq$5.97 Å ($\simeq$304 $\mathrm{km\,s^{-1}}$) apart, a medium-resolution resolving power of $\lambda/\Delta\lambda \gtrsim 10^3$ suffices. Making this advantage quantitative and demonstrating that essentially no accuracy is sacrificed outside the strongly saturated regime is a central result of this paper.

We show that the classical Gaussian COG inversion can be reformulated as a saturation correction that depends only on the observed doublet ratio, reducing the two-parameter inversion to a one-dimensional problem. This formulation naturally yields both an exact inversion for a single Gaussian absorber, $N_{\tilde{\kappa}}$, and a simple closed-form hierarchy

$$N_{\rm thin} \le N_{\rm 2nd} \le N_{\rm min} \le N_{\rm true}, \qquad (2)$$

where $N_{\rm thin}$ is the optically thin estimate, $N_{\rm 2nd}$ is the second-order approximation, $N_{\rm min}$ is an optimal strict bound attained by a square optical-depth profile, and $N_{\rm true}$ is the true line-of-sight column density; the chain holds for an arbitrary line-of-sight velocity structure. We recommend $N_{\rm min}$ as the reported column density: It requires no assumptions about the velocity structure of the absorber, and for $R \gtrsim 1.4$ it agrees to better than a few per cent with $N_{\tilde{\kappa}}$. The exact Gaussian inversion remains the model-based estimate and the only route to the Doppler parameter; it exceeds $N_{\rm min}$ by at most a known, bounded factor $\eta(R)$. We derive each estimator, establish the origin of the successive inequalities, and examine their accuracy for both single- and multicomponent absorption systems. A practical consequence of the hierarchy is that the ratio $N_{\tilde{\kappa}}/N_{\rm 2nd}$ measures the strength of the saturation correction, providing a built-in diagnostic of the reliability of the inferred column density for each individual object.

A primary motivation for this formalism is the determination of dust extinction. P13 found a tight correlation between the Na I column density and the visual extinction $A_V$ along Milky Way lines of sight, whereas calibrations based on the raw Na ID EW (L. M. Hobbs 1974; U. Munari & T. Zwitter 1997; B. Y. Welsh et al. 2010; D. Poznanski et al. 2011; D. Poznanski, J. X. Prochaska & J. S. Bloom 2012) suffer from the degeneracy between column density and velocity structure, and have been shown to depend on the host-galaxy environment and the underlying population (S. González-Gaitán et al. 2024, 2025, 2026; C. P. Gutiérrez et al. 2026). The formalism presented here removes the main obstacle to using the $\log N(\text{Na I})$–$A_V$ relation at scale: It converts two EW measurements, obtainable from medium-resolution spectra, into a column density with a controlled saturation correction. This opens a route to extinction estimates for large samples of Galactic and extragalactic objects – in particular, transients, for which a reliable extinction-free reference for the intrinsic emission is typically unavailable. In this context, Type Ia supernovae (SNe Ia) are a natural first application because their Na ID absorption is a well-studied observable: It has long served as an empirical dust-extinction proxy (D. Poznanski et al. 2012, P13), and its velocity structure and temporal evolution have been extensively investigated as possible signatures of circumstellar material (F. Patat et al. 2007; J. D. Simon et al. 2009; A. Sternberg et al. 2011; K. Maguire et al. 2013; N. Soker 2014; A. Sternberg et al. 2014; P. Clark et al. 2021; C. Gall et al. 2024), so suitable spectra already exist in large numbers. Moreover, SNe Ia offer an independent check on any extinction estimate by using the well-defined blue edge of the observed colour distribution (C. R. Burns et al. 2014, 2018). We therefore test the formalism against independent profile-fitting Na I column densities along both Galactic sightlines and sightlines towards SNe Ia. We emphasize that the extinction calibration enters only at the final conversion step: The formalism yields $N$(Na I) from the two measured EWs alone, so a revised or alternative $\log N$(Na I)–$A_V$ relation can be adopted directly, without any change to the method or to the inferred column densities.

The remainder of this paper is organized as follows. Section 2 develops the theoretical framework for a general doublet: the one-dimensional inversion together with the scaling symmetry and the line-profile conditions underlying it, the domain of validity of the Gaussian profile, the closed-form second-order approximation, and the behaviour for multicomponent lines of sight, including the strict lower bound $N_{\rm min}$, which we prove is attained by a box-shaped (square) optical-depth profile. Section 3 applies the formalism to the Na ID doublet: It presents the inversion in practice and validates the method against published profile-fitting measurements of Galactic sightlines (Section 3.2) and SNe Ia (Section 3.3). Section 4 summarizes the results and discusses their implications.

## 2 THEORY

In this section, we develop the theoretical framework for a general resonance-line doublet. In Section 2.1, we show that the classical Gaussian COG inversion can be reformulated as a universal saturation correction depending only on the observed doublet ratio, reducing the classical two-parameter inversion to a one-dimensional problem; we identify the scaling symmetry responsible for this structure and specify the conditions that the line profile must satisfy for the analysis to apply, and delineate the domain of validity of the underlying Gaussian line profile. In Section 2.2, we derive a simple closed-form second-order approximation and establish the resulting hierarchy of lower bounds. In Section 2.3, we examine the behaviour of the estimators for multicomponent lines of sight; there, we also introduce the strict lower bound $N_{\rm min}$ and prove that it is attained by a box-shaped profile. The propagation of the EW measurement uncertainties to the inferred column densities is discussed in Appendix A.

Throughout, we label the two members of the doublet such that transition 2 is the stronger one, $f_2\lambda_2 > f_1\lambda_1$, where $f_i$ and $\lambda_i$ are the oscillator strengths and rest wavelengths.

### 2.1 The doublet inversion is one-dimensional

We begin by considering a Gaussian absorption doublet with measured $\mathrm{EW}_2$ and $\mathrm{EW}_1$. Throughout this subsection, the absorber is a single-velocity component with a Gaussian optical-depth profile; lines of sight containing multiple components are treated in Section 2.3. In the standard COG formalism, recovering the column density requires solving a coupled non-linear system for the column density $N$ and Doppler parameter $b$. We show that, for a Gaussian line profile, the exact COG inversion possesses a remarkably simple structure:

$$N_{\rm exact} = \mathrm{EW}_2\,\kappa(R), \tag{3}$$

where $R$ is the doublet ratio defined in equation (1) and $\kappa(R)$ is a function of $R$ alone. Thus, at a fixed doublet ratio, the exact column density is linear in the EW of one of the transitions, while the dependence on the Doppler parameter is fully encoded in the observed ratio.

We begin from the standard COG expression for the EW of a Doppler-broadened absorption line,

$$\mathrm{EW} = \frac{\lambda}{c}\, b\,\Phi(\tau_0), \tag{4}$$

where

$$\Phi(\tau_0) = \int_{-\infty}^{+\infty} \left[1 - e^{-\tau_0 e^{-u^2}}\right] \mathrm{d}u, \tag{5}$$

$u = (v - v_0)/b$ is the dimensionless velocity relative to the line centre, and $\tau_0$ is the optical depth at line centre:

$$\tau_0 = C_{\rm ph}\, f\lambda\, \frac{N}{b}, \tag{6}$$

with

$$C_{\rm ph} = \sqrt{\pi}\,\frac{e^2}{m_{\rm e} c} = \sqrt{\pi}\, r_0 c, \tag{7}$$

where $e$ is the elementary charge, $m_{\rm e}$ is the electron mass, and $r_0 \equiv e^2/(m_{\rm e}c^2)$ is the classical electron radius.

For an absorption doublet, the two transitions share the same column density and Doppler parameter, giving

$$\mathrm{EW}_2 = \tfrac{\lambda_2}{c}\, b\,\Phi(\tau_{02}), \tag{8}$$
$$\mathrm{EW}_1 = \tfrac{\lambda_1}{c}\, b\,\Phi(\tau_{01}), \tag{9}$$

and the ratio of the line-centre optical depths

$$\frac{\tau_{02}}{\tau_{01}} = \frac{f_2\lambda_2}{f_1\lambda_1} \equiv \mu. \tag{10}$$

Taking the ratio of the two equivalent widths,

$$R = \frac{\mathrm{EW}_2}{\mathrm{EW}_1} = \frac{\lambda_2}{\lambda_1}\frac{\Phi(\tau_{02})}{\Phi(\tau_{01})} = \frac{\lambda_2}{\lambda_1}\frac{\Phi(\tau_{02})}{\Phi(\tau_{02}/\mu)}, \tag{11}$$

the Doppler parameter cancels out. Since $\Phi(\tau_0)$ is a monotonic function, equation (11) uniquely determines $\tau_{02}$ for any measured value of $R$. In the optically thin limit, $\tau_{02} \to 0$, the doublet ratio approaches its maximal value, $R_{\rm thin} = f_2\lambda_2^2/(f_1\lambda_1^2)$ ($\simeq$2.00 for Na ID), while in the strongly saturated limit, $\tau_{02} \to \infty$, both lines lie on the flat part of the COG and $R \to \lambda_2/\lambda_1 \approx 1$. Equation (8) can therefore be solved directly for the Doppler parameter,

$$b = \frac{c\,\mathrm{EW}_2}{\lambda_2\,\Phi(\tau_{02}(R))}. \tag{12}$$

Thus, for a fixed doublet ratio, the Doppler parameter is directly proportional to the EW of one of the transitions.

Using equation (6) and substituting equation (12)

$$N = \frac{\tau_{02}(R)\, b}{C_{\rm ph}\, f_2\lambda_2} = \mathrm{EW}_2\, \frac{c\,\tau_{02}(R)}{C_{\rm ph}\, f_2\lambda_2^2\,\Phi(\tau_{02}(R))}. \tag{13}$$

Defining

$$\kappa(R) \equiv \frac{c\,\tau_{02}(R)}{C_{\rm ph}\, f_2\lambda_2^2\,\Phi(\tau_{02}(R))}, \tag{14}$$

we obtain equation (3).

The function $\kappa(R)$ carries the dimensions of column density per unit equivalent width. To make it dimensionless, we normalize by the single-line optically thin coefficient

$$A_i = \pi r_0 f_i \lambda_i^2, \tag{15}$$

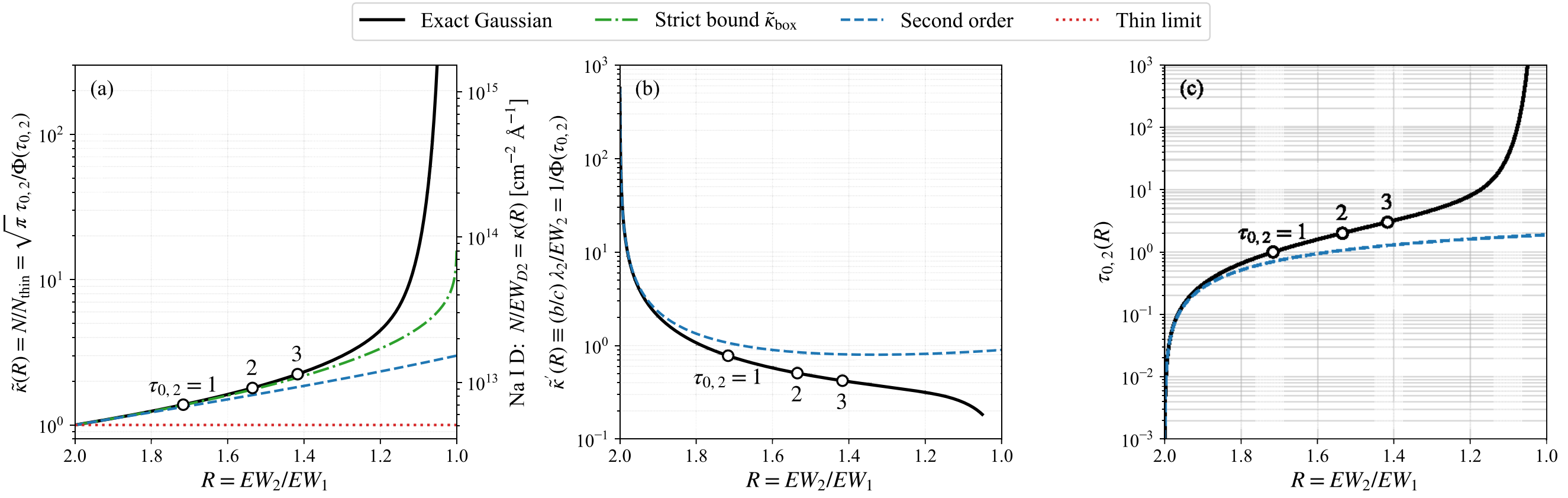

**Figure 1.** The one-dimensional doublet inversion, computed for a doublet with line-centre optical-depth ratio $\mu = f_2\lambda_2/(f_1\lambda_1) = 2$ and $\lambda_1 = \lambda_2$ – an idealization of Na ID to better than 0.1 per cent (for Na ID the induced error in the corrections is a few tenths of a per cent for $R \gtrsim 1.3$, reaching 3 per cent in $\tilde{\kappa}$ at $R = 1.1$; see the text). All panels depend only on the observed doublet ratio, $R = \mathrm{EW}_2/\mathrm{EW}_1$, and are independent of the Doppler parameter. The exact Gaussian COG inversion and the closed-form second-order approximation are shown as solid and dashed curves, respectively; the optically thin limit is shown as a dotted curve, and labelled circles mark $\tau_{0,2} = 1$, 2, and 3 on the exact Gaussian curves. (a) The dimensionless saturation correction, $\tilde{\kappa}(R) = N/N_{\rm thin}$ (equation 18), which converts the optically thin estimate into the exact column density; the right-hand axis gives the corresponding column density per unit equivalent width for the Na ID doublet, $N/\mathrm{EW}_{D2} = \kappa(R)$ (Section 3.1); the dash-dotted curve is the strict lower-bound correction $\tilde{\kappa}_{\rm box}(R)$ (equation 36), valid for an arbitrary velocity structure. (b) The companion function $\tilde{\kappa}'(R)$ (equation 19), which determines the Doppler parameter. (c) The line-centre optical depth $\tau_{0,2}(R)$, the unique solution of equation (11) for each observed doublet ratio. The optically thin approximation is recovered for $R \to R_{\rm thin} = 2$, while the second-order estimator closely follows the exact Gaussian solution up to $\tau_{0,2} \approx 1$ before deviating for strongly saturated absorption.

which relates the EW to the column density in the optically thin limit, $\mathrm{EW}_i = A_i N$. Defining the optically thin estimator

$$N^{\rm thin} \equiv \frac{\mathrm{EW}_2}{A_2}, \qquad (16)$$

the exact column density can be written as

$$N_{\rm exact} = \frac{\mathrm{EW}_2}{A_2}\,\tilde{\kappa}(R), \qquad (17)$$

with the dimensionless universal saturation correction

$$\tilde{\kappa}(R) \equiv A_2\,\kappa(R) = \frac{N_{\rm exact}}{N^{\rm thin}} = \sqrt{\pi}\,\frac{\tau_{02}(R)}{\Phi(\tau_{02}(R))}. \qquad (18)$$

For a fixed observed doublet ratio, the exact column density is thus a linear function of the EW of one of the transitions. Similarly, equation (12) defines a companion correction for the Doppler parameter,

$$\tilde{\kappa}'(R) \equiv \frac{1}{\Phi(\tau_{02}(R))}, \qquad b = \frac{c\,\mathrm{EW}_2}{\lambda_2}\,\tilde{\kappa}'(R), \qquad (19)$$

so that the full two-parameter inversion $(N, b)$ is encoded in the two universal functions $\tilde{\kappa}(R)$ and $\tilde{\kappa}'(R)$.

Fig. 1 presents these functions, computed for a doublet with $\mu = 2$ and $\lambda_1 = \lambda_2$ – an idealization of Na ID , whose parameters equal these values to better than 0.1 per cent. For Ca II H&K, the idealization is somewhat poorer ($\mu \simeq 1.994$ with the D. C. Morton 2003 $f$-values, and $\lambda_{\rm H}/\lambda_{\rm K} = 1.0088$), so its exact parameters should be used in the inversion; the accompanying code accepts a general $(\mu, \lambda_1/\lambda_2)$. Panel (a) shows the saturation correction $\tilde{\kappa}(R)$, which equals unity in the optically thin limit, $R \to R_{\rm thin}$, and rises steeply as $R \to 1$. Panel (c) shows the line-centre optical depth $\tau_{0,2}(R)$, the unique solution of equation (11) for each observed doublet ratio. The circles mark $\tau_{0,2} = 1$, 2, and 3, reached at $R \simeq 1.72$, 1.53, and 1.42, respectively; already at $R = 1.1$, the optical depth is $\tau_{0,2} \simeq 37$. This defines the usable range of the method, $R \gtrsim 1.1$: At smaller ratios, the saturation correction is so steep that realistic EW uncertainties no longer permit a useful column-density estimate (Appendix A). Panel (b) shows the companion function $\tilde{\kappa}'(R)$, which determines the Doppler parameter; the right-hand axis of panel (a) specializes the inversion to the Na ID doublet (Section 3.1).

The green dash–dotted curve in panel (a) anticipates the central practical result of this paper, stated here without proof (the proof is given in Section 2.3.2): For an *arbitrary* line-of-sight velocity structure – any number of components, of any shapes and any degree of overlap – the true column density is bounded from below by $N_{\rm min} \equiv N^{\rm thin}\,\tilde{\kappa}_{\rm box}(R)$, where $\tilde{\kappa}_{\rm box}(R) = \tau_{\rm box}/\left(1 - e^{-\tau_{\rm box}}\right)$ is the saturation correction of a square optical-depth profile and $\tau_{\rm box}(R)$ its depth (equation 36). The bound is optimal – no stronger universal bound exists – and, as the figure shows, it tracks the exact Gaussian inversion to within a few per cent for $R \gtrsim 1.4$ (i.e. $\tau_{0,2} \lesssim 3$), while remaining a valid bound throughout the usable range $R \gtrsim 1.1$ ($\tau_{0,2} \lesssim 37$). This is why, we recommend $N_{\rm min}$ as the reported column density.

The one-dimensional structure of the inversion is a consequence of a scaling symmetry of the COG. Consider rescaling the column density and the Doppler parameter together,

$$N \to \alpha N, \qquad b \to \alpha b, \qquad (20)$$

which corresponds to stretching the absorber's velocity coordinate by a factor $\alpha$ at fixed column density per unit velocity. The line-centre optical depths, $\tau_{0,i} \propto N/b$ (equation 6), are invariant under this transformation, so the doublet ratio is unchanged, while the equivalent widths of both transitions, $\mathrm{EW}_i = (\lambda_i/c)\,b\,\Phi(\tau_{0,i})$, scale by the same factor $\alpha$ as $N$ and $b$. Conversely, since the observed doublet ratio uniquely determines $\tau_{02}$ (equation 11), all absorbers with the same observed doublet ratio form a single one-parameter family, whose members differ only by a common rescaling of EW, $N$, and $b$. At fixed $R$, both $N$ and $b$ must then be linear in the measured EW, with proportionality coefficients that are functions of $R$ alone – precisely the content of equations (17) and (19).

This argument also delineates the conditions that the line profile must satisfy for the doublet-ratio analysis to apply. Nothing in the derivation is specific to the Gaussian shape: It requires only that the optical-depth profile belong to a self-similar, one-parameter family,

$$\tau(v) = \tau_0\,\varphi\left(\frac{v - v_0}{b}\right), \tag{21}$$

with a fixed shape function $\varphi$ (normalized to $\varphi(0) = 1$) and a single width parameter $b$, so that $\tau_0 \propto N/b$. For any such family, equation (4) holds with $\Phi(\tau_0) = \int[1 - e^{-\tau_0\varphi(u)}]\mathrm{d}u$, which is monotonic in $\tau_0$, and the construction of $\tilde{\kappa}(R)$ and $\tilde{\kappa}'(R)$ carries over unchanged, with the specific form of the saturation correction determined by $\varphi$. Broadening mechanisms that introduce a second shape parameter break this self-similarity; the most important example is the damping wings of the Voigt profile. Their importance is controlled by the Voigt parameter $a = \Gamma\lambda/(4\pi b)$, where $\Gamma$ is the radiative damping rate of the transition: the wings dominate the growth of the EW once $a\tau_0/\sqrt{\pi} \gtrsim \ln\tau_0$. For optical resonance lines $a \sim 10^{-3}$ to $10^{-2}$ [for the Na ID and Ca II H&K doublets, $a \approx 3$–$5 \times 10^{-3}\,(b/1\,\mathrm{km\,s^{-1}})^{-1}$; D. C. Morton 2003], so the Gaussian COG is accurate up to line-centre optical depths of order $10^3$ to $10^4$ – roughly two orders of magnitude beyond the $\tau_0$ of a few tens at which the doublet ratio saturates ($R \to 1$) and loses its sensitivity to the column density.[1] We therefore adopt the Gaussian shape – thermal and turbulent broadening dominate interstellar absorption lines wherever the doublet ratio carries usable information – and the formalism is not intended for the damping-dominated regime.

### 2.2 Closed-form second-order approximation

Expanding the integrand in a Taylor series in $\tau_0$ of equation (5) and integrating term by term yields:

$$\Phi(\tau_0) = \sqrt{\pi}\sum_{n=1}^{\infty}\frac{(-1)^{n+1}\,\tau_0^n}{n!\,\sqrt{n}} = \sqrt{\pi}\,\tau_0 - \frac{\sqrt{\pi}}{2\sqrt{2}}\,\tau_0^2 + \frac{\sqrt{\pi}}{6\sqrt{3}}\,\tau_0^3 - \cdots. \tag{22}$$

Truncating equation (22) at second order gives the workhorse expression

$$\mathrm{EW}_i \;=\; A_i N \;-\; \frac{B_i}{b}N^2 \;+\; \mathcal{O}(N^3/b^2), \tag{23}$$

where $B_i = \frac{\pi^{3/2}}{2\sqrt{2}}\,r_0^2\,c\,f_i^2\,\lambda_i^3$. Dividing by $A_i$ and using the definition of $N_{\mathrm{thin}}$ from equation (16),

$$N_{\mathrm{thin},i} = \frac{\mathrm{EW}_i}{A_i} \;=\; N - \beta_i\,\frac{N^2}{b}, \tag{24}$$

where $\beta_i \;\equiv\; \frac{B_i}{A_i} = \frac{\sqrt{\pi}}{2\sqrt{2}}\,r_0\,c\,f_i\,\lambda_i$.

The two equations for the two transitions of the doublet are linear in $(N,\ N^2/b)$ and have the unique solution

$$N_{\mathrm{2nd}} \;=\; \frac{\beta_2\,N_{\mathrm{thin},1} - \beta_1\,N_{\mathrm{thin},2}}{\beta_2 - \beta_1}, \tag{25}$$
$$b_{\mathrm{eff}}^{\mathrm{2nd}} \;=\; \frac{N_{\mathrm{2nd}}^2\,(\beta_2 - \beta_1)}{N_{\mathrm{thin},1} - N_{\mathrm{thin},2}}.$$

Substituting the definition of $\beta_i$ gives the equivalent, doublet-independent form

$$N_{\mathrm{2nd}} = \frac{f_2\lambda_2\,N_{\mathrm{thin},1} - f_1\lambda_1\,N_{\mathrm{thin},2}}{f_2\lambda_2 - f_1\lambda_1}. \tag{26}$$

Equation (26) is algebraically akin to the doublet correction of B. D. Savage & K. R. Sembach (1991), which is likewise built from the difference of the two apparent (optically thin) column densities; here, the combination emerges as the exact quadratic-order inversion. Dividing by $N_{\mathrm{thin},2} = \mathrm{EW}_2/A_2$ recasts the solution as the second-order saturation correction

$$\tilde{\kappa}_{\mathrm{2nd}}(R) \;=\; \frac{\beta_2\,A_2/(R\,A_1) - \beta_1}{\beta_2 - \beta_1}, \tag{27}$$

Since $1 - e^{-\mu\tau} \leq \mu\left(1 - e^{-\tau}\right)$ pointwise, every true absorber satisfies $R \leq R_{\mathrm{thin}}$, and equation (27) then gives $\tilde{\kappa}_{\mathrm{2nd}}(R) \geq 1$: The second-order estimate never falls below the optically thin one, $N^{\mathrm{thin}} \leq N_{\mathrm{2nd}}$. Equation (27) is a hyperbola in $R$ that is identically the universal recipe equation (17) with $\Phi$ replaced by its second-order Taylor truncation.

The second-order system also yields the companion correction for the Doppler parameter: In analogy with equation (19), $b_{\mathrm{eff}}^{\mathrm{2nd}} = (c\,\mathrm{EW}_2/\lambda_2)\,\tilde{\kappa}'_{\mathrm{2nd}}(R)$, with $\tilde{\kappa}'_{\mathrm{2nd}}(R) = 1/\Phi_{\mathrm{2nd}}\left(\tau_{0,2}^{\mathrm{2nd}}(R)\right)$, where $\Phi_{\mathrm{2nd}}$ is the quadratic truncation of equation (22) and $\tau_{0,2}^{\mathrm{2nd}}(R)$ solves equation (11) with $\Phi \to \Phi_{\mathrm{2nd}}$. Fig. 1 compares the second-order and exact Gaussian inversions: All three functions track the exact Gaussian solution closely up to $\tau_{0,2} \approx 1$, corresponding to $R \gtrsim 1.7$ (panel c), beyond which the truncation returns a smaller optical depth, and hence a smaller column density. The Doppler parameter behaves oppositely (panel b): Because the truncation underestimates the saturation correction, it attributes the same equivalent widths to a less saturated but broader line, so that $\tilde{\kappa}'_{\mathrm{2nd}} \geq \tilde{\kappa}'$ and the second-order estimate of $b$ exceeds the true value, mirroring the underestimate of the column density.

Higher-order Taylor approximations can be derived straightforwardly, but they no longer admit comparably simple closed-form solutions. More importantly, beginning at the third order, the expansion depends on higher moments of the column-density distribution that cannot be represented by a single effective Doppler parameter. Consequently, the multicomponent blindness of the second-order estimator, derived next, is unique to quadratic order. While higher-order truncations improve the optically thin regime, they offer little benefit in saturated systems and sacrifice the simplicity and robustness of the second-order approximation.

### 2.3 Multicomponent lines of sight

Real lines of sight often contain multiple absorbing components, even within the interstellar medium of the Milky Way (see e.g. figs 1, 2, and 7 of P13). In this subsection, we examine how the two estimators behave for such systems: the closed-form second-order estimator (Section 2.3.1) and the exact Gaussian inversion (Section 2.3.2). The latter culminates in the derivation of the distribution-free strict bound $N_{\mathrm{min}}$.

[1]For Na ID, the $D_2$ damping rate $\Gamma \simeq 6.16 \times 10^7\,\mathrm{s^{-1}}$ (D. C. Morton 2003) gives $a \simeq 2.9 \times 10^{-3}\,(b/1\,\mathrm{km\,s^{-1}})^{-1}$, so the wings dominate only at $\tau_{0,\mathrm{D_2}} \gtrsim 5 \times 10^3$ for $b = 1\,\mathrm{km\,s^{-1}}$ (larger for broader lines), i.e. $N(\mathrm{Na\,I}) \gtrsim 10^{15}\,\mathrm{cm^{-2}}$, whereas the doublet ratio is informative only for $\tau_{0,\mathrm{D_2}} \lesssim 40$ ($R \gtrsim 1.1$; Fig. 1).

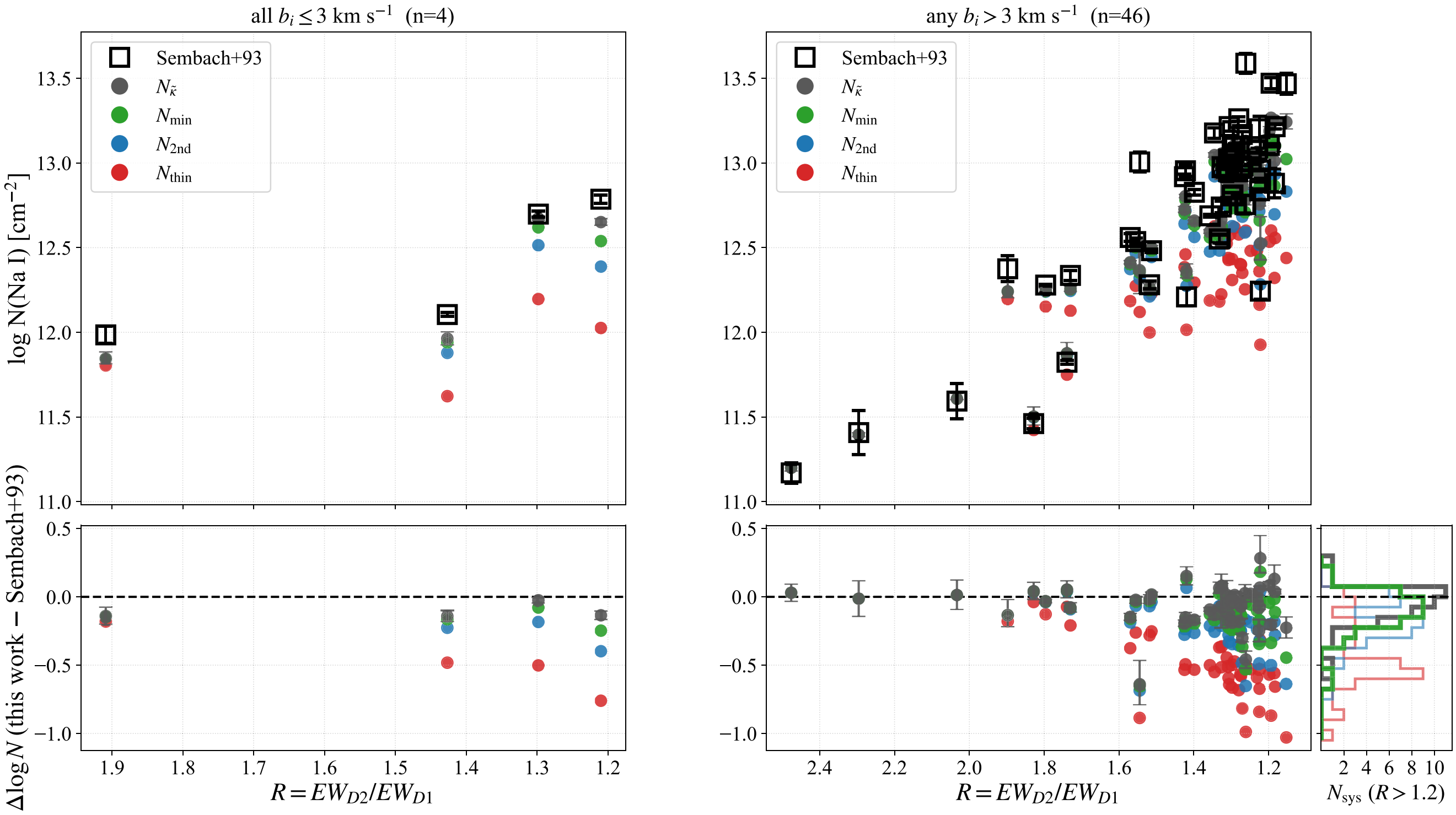

**Figure 2.** Comparison of Na I column densities inferred for the Galactic sightlines of K. R. Sembach et al. (1993) using the four estimators discussed in this work, separated according to the Doppler parameters of the individual components obtained from their profile-fitting analysis. Left: Sightlines for which all fitted components have $b_i \leq 3\,\mathrm{km\,s^{-1}}$ ($n = 4$). Right: Sightlines containing at least one component with $b_i > 3\,\mathrm{km\,s^{-1}}$ ($n = 46$). Top: Column densities derived from the exact Gaussian inversion $N_{\tilde{\kappa}}$ (grey), the recommended strict bound $N_{\rm min}$ (green), the closed-form second-order estimator $N_{\rm 2nd}$ (blue), and the optically thin approximation $N^{\rm thin}$ (red), compared with the total column densities reported by K. R. Sembach et al. (1993) (open black squares); the estimator colours follow the curves of Fig. 1. Bottom: Difference between each estimate and the published profile-fitting column density as a function of the observed doublet ratio $R$; the error bars combine our uncertainties with the published ones in quadrature. For clarity, error bars are shown only for $N_{\tilde{\kappa}}$ and for the published values; our uncertainties represent the 16th–84th percentile confidence intervals obtained by Monte Carlo propagation of the measured equivalent-width uncertainties. All estimators converge in the optically thin regime ($R \approx 2$), while their differences increase towards smaller $R$ as the absorption becomes increasingly saturated, preserving the hierarchy of equation (2). Bottom right: Histograms of the residuals for the systems with informative doublet ratios ($R > 1.2$), shown in the same colours.

### 2.3.1 *The second-order estimator*

The crucial property of $\tilde{\kappa}_{\rm 2nd}$ is that it is insensitive to the velocity structure of the absorbing gas. Consider first a multicloud LOS with $n$ components $(N_j, b_j)$ whose velocity separations are large compared with their widths, so that the absorption troughs do not overlap and the total equivalent widths are sums,

$$\mathrm{EW}_i^{\rm tot} = \sum_{j=1}^{n} \frac{\lambda_i b_j}{c}\,\Phi\left(\tau_{0,i}^{(j)}\right), \qquad \tau_{0,i}^{(j)} = \frac{N_j \sqrt{\pi}\, r_0\, c\, f_i\, \lambda_i}{b_j}. \tag{28}$$

At second order, the sum becomes $\mathrm{EW}_i^{\rm tot} = A_i \sum_j N_j - B_i \sum_j N_j^2/b_j$. Demanding it be reproduced by a single equivalent Gaussian $(N_{\rm eff}, b_{\rm eff})$ with $N_{\rm eff} = \sum_j N_j$ leaves one equation per line,

$$B_i \frac{(\sum_j N_j)^2}{b_{\rm eff}} = B_i \sum_j \frac{N_j^2}{b_j}. \tag{29}$$

The line-dependent factor $B_i$ cancels uniformly between the two transitions, leaving a single equation with a single solution

$$b_{\rm eff} = \frac{\left(\sum_j N_j\right)^2}{\sum_j N_j^2/b_j}, \tag{30}$$

that satisfies both lines simultaneously. Thus, to second order, every multicomponent LOS is observationally indistinguishable from a single Gaussian with column $\sum_j N_j$ and Doppler parameter $b_{\rm eff}$ of equation (30). The closed-form equation (27) therefore returns the true $\sum_j N_j$ exactly whenever the second-order Taylor expansion of $\Phi$ is accurate, regardless of how the column is distributed in velocity space (still for non-overlapping components; overlap is treated below). Two clean limits: $n$ identical clouds give $b_{\rm eff} = n b_0 = \sum_j b_j$; one cloud dominating gives $b_{\rm eff} \to b_1$.

The derivation above assumes that the components do not overlap in velocity, so that equation (28) applies. When components overlap, the exponentials no longer add: The total optical depth is $\tau_i(v) = \sum_j \tau_i^{(j)}(v)$, and the blended EW is smaller than the sum of the individual ones. At second order, however, the blindness persists. Expanding $1 - \exp[-\tau_i(v)]$ to quadratic order, the term $\propto \int \tau_i(v)^2\, dv$ now contains cross terms between overlapping components; since $\tau_i^{(j)} \propto f_i\lambda_i$ for every component, these cross terms carry the same line-index factor $B_i \propto f_i^2\lambda_i^3$ as the diagonal ones, and cancel between the two transitions in exactly the same way. The only effect of overlap is to generalize the effective Doppler parameter of equation (30) to

$$\frac{\left(\sum_j N_j\right)^2}{b_{\rm eff}} = \sqrt{2}\sum_{j,k}\frac{N_jN_k}{\sqrt{b_j^2+b_k^2}}\exp\left(-\frac{\Delta v_{jk}^2}{b_j^2+b_k^2}\right), \quad (31)$$

where $\Delta v_{jk}$ is the velocity separation between the centres of components $j$ and $k$. For well-separated components ($|\Delta v_{jk}| \gg b_j, b_k$) the off-diagonal terms vanish and equation (30) is recovered, while for fully overlapping identical components equation (31) reduces to the single-cloud result. The second-order estimator therefore returns the true $\sum_j N_j$ whenever the quadratic truncation is accurate for the blended profile, regardless of both the velocity distribution and the degree of overlap of the components.

$N_{\rm 2nd}$ provides a lower bound on the true column density for an arbitrary velocity structure. For separated components, this follows from additivity: Because $N_{\rm 2nd}$ is linear in the equivalent widths,

$$N_{\rm 2nd} \;=\; \sum_j N_{\rm 2nd}^{(j)} \;\le\; \sum_j N_j, \quad (32)$$

where the inequality follows because each individual Gaussian component satisfies $N_{\rm 2nd}^{(j)} \le N_j$ (the second-order truncation underestimates the saturation correction). The bound, however, relies neither on the components being separated nor on the profile being Gaussian. For an arbitrary optical-depth profile $\tau_1(v)$ of the weak transition (the strong transition then has $\mu\,\tau_1(v)$), write the integrand of the EW as $1-e^{-\tau} = \tau - \tau^2/2 + g(\tau)$, where $g(\tau) \equiv 1 - e^{-\tau} - \tau + \tau^2/2 \ge 0$ is the truncation remainder. The linear term integrates to $\pi r_0 c f_i \lambda_i N_{\rm true}$ exactly, independent of the velocity structure, and the quadratic terms cancel in the combination of equation (26), leaving

$$N_{\rm true} - N_{\rm 2nd} = \frac{\int\left[g(\mu\,\tau_1(v)) - \mu^2 g(\tau_1(v))\right]{\rm d}v}{\pi r_0 c f_1 \lambda_1\,\mu\,(\mu-1)} \;\ge\; 0, \quad (33)$$

where the inequality holds pointwise, because $g(x)/x^2$ is monotonically increasing and $\mu > 1$. The lower-bound property of $N_{\rm 2nd}$ therefore holds for any number of components, any degree of overlap, and any line profile, with equality approached only when every component lies within the regime where the second-order Taylor approximation is exact.

### 2.3.2 *The exact Gaussian inversion and the strict bound $N_{\rm min}$*

For a multicomponent line of sight, the summed EWs (equation 28) are, in general, not representable by a single Gaussian profile. Consequently, the exact Gaussian inversion recovers the unique equivalent Gaussian that reproduces the observed doublet equivalent widths. An important exception occurs when all (non-overlapping) components satisfy $N_1/b_1 = N_2/b_2 = \cdots$, i.e. have the same line-centre optical depth. In this case, the scaling relation derived in Section 2.1 applies independently to each component, and the summed EWs are exactly reproduced by a single equivalent Gaussian with

$$N_{\tilde{\kappa}} = \sum_j N_j, \qquad b_{\rm eff} = \sum_j b_j. \quad (34)$$

Thus, the exact Gaussian inversion recovers the true total column density along the equal-$\tau$ locus. Away from this locus, the summed EWs of a blend are in general not those of any single Gaussian, and the exact Gaussian inversion misestimates the true column density by an amount that depends on the unknown velocity structure. Rather than cataloging configurations, we ask the worst-case question directly; its answer is the strict bound $N_{\rm min}$ anticipated in Section 2.1.

Consider the smallest true column density consistent with the measured (${\rm EW}_1$, ${\rm EW}_2$), over *all* optical-depth profiles $\tau_2(v) \ge 0$ – any number of components, any degree of overlap, and any component shapes. Writing $w(\tau)\,{\rm d}\tau$ for the total velocity interval occupied by optical depths in $[\tau, \tau + {\rm d}\tau]$, the two measurements and the column density, ${\rm EW}_2 \propto \int(1-e^{-\tau})\,w\,{\rm d}\tau$, ${\rm EW}_1 \propto \int(1 - e^{-\tau/\mu})\,w\,{\rm d}\tau$, and $N \propto \int \tau\, w\,{\rm d}\tau$, are all linear in $w$.

Minimizing $N$ over $w \ge 0$ at fixed (${\rm EW}_1$, ${\rm EW}_2$) is a linear programme, but its solution can be established directly, with no optimization machinery. Let $\tau_{\rm box}$ denote the depth of the unique single-level (square) profile that reproduces the two measured EWs: Its value is fixed by the observed doublet ratio (equation 36 below), and the profile's width is set by the overall EW scale. With this depth in hand, define the auxiliary function $h(\tau) \equiv \tau - c_1\left(1-e^{-\tau}\right) - c_2\left(1-e^{-\tau/\mu}\right)$, where the two constants ($c_1$, $c_2$) are fixed by requiring $h$ to be tangent to zero at this depth: $h(\tau_{\rm box}) = 0$ and $h'(\tau_{\rm box}) = 0$, two linear equations for the two constants. Being determined by the measurements alone, ($c_1$, $c_2$) are known numbers; they are independent of the unknown profile. Because ${\rm EW}_1$, ${\rm EW}_2$, and $N$ are all linear in $w$, every profile consistent with the measurements obeys the identity

$$\int \tau\, w\,{\rm d}\tau = \int h(\tau)\,w\,{\rm d}\tau + c_1\int\left(1-e^{-\tau}\right)w\,{\rm d}\tau + c_2\int\left(1-e^{-\tau/\mu}\right)w\,{\rm d}\tau, \quad (35)$$

in which the last two integrals are fixed by the measured ${\rm EW}_2$ and ${\rm EW}_1$. Both sides of equation (35) refer to one and the same profile $w$. The left-hand side is proportional to that profile's column density. On the right-hand side, the values of the two EW integrals are set by the measurements, and ($c_1$, $c_2$) are constants: The last two terms therefore take identical values for every profile consistent with the data. Any two admissible profiles thus differ in column density only through the first term, $\int h(\tau)\,w\,{\rm d}\tau$. Everything now rests on a single property of the specific $h$ defined above, established in the next paragraph: $h(\tau) \ge 0$ for all $\tau \ge 0$, with equality only at $\tau = 0$ and at $\tau = \tau_{\rm box}$. Granting it, the term $\int h(\tau)\,w\,{\rm d}\tau$ is non-negative for every admissible $w \ge 0$, and it vanishes precisely for profiles whose weight is confined to $\tau = 0$ and $\tau = \tau_{\rm box}$ – the square profile among them.

The function $h$ is non-negative everywhere, by three elementary observations. First, $h(0) = 0$ and $h(\tau \to \infty) \to \infty$. Secondly, $h'(\tau) = 1 - c_1 e^{-\tau} - (c_2/\mu)\,e^{-\tau/\mu}$ vanishes at most twice: The substitution $x = e^{-\tau/\mu}$ turns it into $1 - c_1 x^\mu - (c_2/\mu)x$, whose curvature in $x$ has a single sign, and such a function has at most two roots, counting multiplicity. Thirdly, both roots are accounted for: One is the tangency point $\tau_{\rm box}$, and Rolle's theorem places another, $\tau_1$, strictly inside $(0, \tau_{\rm box})$, since $h$ vanishes at both ends of this interval. With the budget of two roots exhausted, both are simple, so $h'$ changes sign at each. Beyond $\tau_{\rm box}$, $h'$ must be positive, because $h$ starts there from zero and diverges; the alternation then forces $h' < 0$ on $(\tau_1, \tau_{\rm box})$ and $h' > 0$ on $(0, \tau_1)$. The shape of $h$ follows: It rises from $h(0) = 0$ to a single maximum at $\tau_1$, descends back to zero at $\tau_{\rm box}$, and increases without bound thereafter. Hence, $h \geq 0$, with equality only at $\tau = 0$ and $\tau = \tau_{\rm box}$.

Every admissible profile therefore has a column density at least that of the square profile, with equality only if its weight is confined to $\tau = 0$ and $\tau = \tau_{\rm box}$; weight at $\tau = 0$ contributes neither absorption nor column density, so the minimizer is the square (top-hat) profile. The resulting strict lower bound is

$$N_{\rm min} = N^{\rm thin}\,\tilde{\kappa}_{\rm box}(R), \qquad \tilde{\kappa}_{\rm box}(R) = \frac{\tau_{\rm box}(R)}{1 - e^{-\tau_{\rm box}(R)}}, \tag{36}$$

where $\tau_{\rm box}(R)$ is the unique solution of $\left(1 - e^{-\tau}\right) / \left(1 - e^{-\tau/\mu}\right) = (\lambda_1/\lambda_2)\,R$. Writing $r \equiv (\lambda_1/\lambda_2)\,R$, two limits follow from expanding the defining equation. In the optically thin limit, $r \to \mu$ (i.e. $R \to R_{\rm thin}$), $\tau_{\rm box} \to 2\,(\mu - r)/(\mu - 1) \to 0$, and $\tilde{\kappa}_{\rm box} \to 1$, recovering $N_{\rm min} \to N^{\rm thin}$; in the strongly saturated limit, $r \to 1$ (i.e. $R \to \lambda_2/\lambda_1$), $\tau_{\rm box} \to \mu \ln\left[1/\left(r - 1\right)\right] \to \infty$, and $\tilde{\kappa}_{\rm box} \simeq \tau_{\rm box}$, so the bound grows only logarithmically as the doublet ratio approaches its saturated value. The bound $N_{\rm min} \leq N_{\rm true}$ holds for an arbitrary velocity structure, and it is optimal in the sense that no stronger universal bound exists: Many narrow, strongly saturated components placed side by side produce a total optical-depth profile that approaches a square, so there exist genuine multi-component absorbers, consistent with the same measured EWs, whose true column density lies arbitrarily close to $N_{\rm min}$ – any prospective universal bound larger than $N_{\rm min}$ would be violated by such absorbers.

The remaining link of equation (2), $N_{\rm 2nd} \leq N_{\rm min}$, follows in three steps. First, once $({\rm EW}_1, {\rm EW}_2)$ are measured, $N_{\rm 2nd}$ is a fixed number (equation 26) – it does not depend on which absorber produced the EWs. Secondly, equation (33) states that $N_{\rm 2nd} \leq N_{\rm true}$ for *every* absorber consistent with these EWs; applying it to the minimizing square absorber, whose true column density is $N_{\rm min}$ by construction, gives $N_{\rm 2nd} \leq N_{\rm min}$. Thirdly, combining this with $N^{\rm thin} \leq N_{\rm 2nd}$ (Section 2.2) and with $N_{\rm min} \leq N_{\rm true}$ ($N_{\rm min}$ is the minimum of $N$ over all profiles reproducing the measured EWs, a set that contains the actual absorber), the chain $N^{\rm thin} \leq N_{\rm 2nd} \leq N_{\rm min} \leq N_{\rm true}$ holds for an arbitrary velocity structure.

The function $\tilde{\kappa}_{\rm box}(R)$ is shown in panel (a) of Fig. 1: It tracks the exact Gaussian correction closely over most of the usable range. Consequently, the maximal factor by which $N_{\tilde{\kappa}}$ can overestimate the true column density is $\eta(R) \equiv \tilde{\kappa}(R)/\tilde{\kappa}_{\rm box}(R)$, the separation between the exact Gaussian and strict-bound curves in panel (a) of Fig. 1 (the values that follow are computed for the idealized doublet of Fig. 1; for the exact Na ID parameters they agree to better than 0.4 per cent at $R \geq 1.2$): $\eta \leq 1.01$ for $R \geq 1.7$, and $\eta \simeq 1.06$, 1.33, and 3.5 (3.4 for the exact Na ID parameters) at $R = 1.4$, 1.2, and 1.1, respectively. No absorber, however contrived, can exceed this factor. For intuition, one concrete case: two overlapping components with $(\tau_{0,2}, b) = (3.5,\ 0.5\,{\rm km\,s^{-1}})$ and $(5.3,\ 1.3\,{\rm km\,s^{-1}})$, separated by $1.9\,{\rm km\,s^{-1}}$, yield $R = 1.23$ and $N_{\tilde{\kappa}}/N_{\rm true} = 1.11$ – a narrow saturated component hiding within a broader one. We therefore recommend $N_{\rm min}$ as the reported column density: It is a certified, optimal lower limit for an arbitrary velocity structure, and since $\eta \leq 1.06$ for $R \gtrsim 1.4$, essentially no accuracy is sacrificed there relative to the exact Gaussian inversion; towards the saturated limit ($\eta = 1.33$ at $R = 1.2$, 3.5 at $R = 1.1$) the bound becomes increasingly conservative, as a distribution-free statement must. $N_{\tilde{\kappa}}$ remains the exact solution under the single-Gaussian model – and the only route to the Doppler parameter (equation 19) – but for the column density it should be regarded as the model-based estimate rather than the primary value.

### 2.4 Summary of the estimators

The analysis of this section provides four estimators of the column density, all computable from the two measured EWs alone. Three of them are ordered by the chain of equation (2), $N^{\rm thin} \leq N_{\rm 2nd} \leq N_{\rm min} \leq N_{\rm true}$, which holds for an arbitrary line-of-sight velocity structure. The optically thin estimate $N^{\rm thin}$ and the closed-form second-order approximation $N_{\rm 2nd}$ require no numerical inversion and become accurate towards the optically thin limit. The recommended value is $N_{\rm min}$: a certified, optimal lower bound on the true column density for an arbitrary velocity structure, obtained from the observed doublet ratio through equation (36). The exact Gaussian inversion is tied to it by $N_{\tilde{\kappa}} = \eta(R)\,N_{\rm min}$, with $\eta$ known – $\eta \leq 1.06$ for $R \gtrsim 1.4$, rising to 3.5 at $R = 1.1$; it is the exact solution under the single-Gaussian model and the only route to the Doppler parameter, but for the column density it should be regarded as the model-based estimate rather than the primary value.

The hierarchy of equation (2) has a useful practical corollary. Since $N_{\rm 2nd} \leq N_{\tilde{\kappa}}$, and since the two estimators coincide in the regime where the second-order expansion is accurate, the ratio $N_{\tilde{\kappa}}/N_{\rm 2nd}$ directly measures the strength of the saturation correction beyond second order. When $N_{\tilde{\kappa}}/N_{\rm 2nd} \approx 1$, the absorption is at most mildly saturated, the sensitivity to unresolved velocity structure is weak, and $N_{\tilde{\kappa}}$ is a reliable estimate of the true column density. This internal consistency check requires no information beyond the two measured EWs, and we recommend reporting it alongside any inferred column density. The ratio $N_{\tilde{\kappa}}/N^{\rm thin} = \tilde{\kappa}(R)$ would instead measure the *total* saturation correction; since its part through second order is captured by $N_{\rm 2nd}$ for any velocity structure, it is the excess of $N_{\tilde{\kappa}}$ over $N_{\rm 2nd}$, rather than over $N^{\rm thin}$, that isolates the model-sensitive part of the correction.

A reference implementation of all inversion functions – $\tilde{\kappa}(R)$, $\tilde{\kappa}'(R)$, $\tau_{0,2}(R)$, their second-order counterparts, and $\tilde{\kappa}_{\rm box}(R)$, for a general doublet – is provided with this paper (see Data Availability). Table 1 lists these functions at several doublet ratios for the $\mu = 2$, $\lambda_1 = \lambda_2$ case, for benchmarking independent implementations.

## 3 APPLICATION TO THE Na ID DOUBLET

We now specialize the formalism to the Na ID doublet, with rest wavelengths $\lambda_{\rm D_2} \simeq 5889.95$ Å and $\lambda_{\rm D_1} \simeq 5895.92$ Å and oscillator strengths $f_{\rm D_2} \simeq 0.641$ and $f_{\rm D_1} \simeq 0.320$ (e.g. D. C. Morton 2003). The line-centre optical-depth ratio is $\mu = f_2\lambda_2/(f_1\lambda_1) \simeq 2.001$, and the optically thin doublet ratio is $R_{\rm thin} = f_2\lambda_2^2/(f_1\lambda_1^2) \simeq 1.999$; both are equal to 2 to better than 0.1 per cent (the last quoted digit is sensitive to the adopted $f$-values at the $\pm 0.001$ level). In this limit, the second-order correction (equation 27) collapses to the compact form

$$\tilde{\kappa}_{\rm 2nd}(R) \xrightarrow{f_2\lambda_2 = 2 f_1\lambda_1} \frac{4}{R} - 1, \tag{37}$$

and, equivalently, $N_{\rm 2nd} \approx 2\,N_{\rm thin,D_1} - N_{\rm thin,D_2}$. The strict bound collapses similarly: For $\mu = 2$, the equation defining $\tau_{\rm box}$ (Section 2.3.2) is a quadratic in $e^{-\tau/2}$ and reduces to $1 + e^{-\tau/2} = r$, where $r \equiv (\lambda_1/\lambda_2)\,R \approx R$ (the wavelength ratio is 1.0010), so that $\tau_{\rm box}(R) \simeq 2\ln\left[1/\left(R - 1\right)\right]$, the wavelength-ratio correction shifting $\tilde{\kappa}_{\rm box}$ by less than 0.5 per cent across the usable range, and, since $e^{-\tau_{\rm box}} = (R - 1)^2$, equation (36) collapses to the fully closed form $\tilde{\kappa}_{\rm box}(R) = 2\ln\left[1/\left(R - 1\right)\right] / \left[1 - (R - 1)^2\right]$.

Two practical facts frame this application. First, since the EW of a line is preserved under convolution with the instrumental line-spread function, the only observational requirement of the method is that the two doublet members be measured separately: For Na ID, whose members are $\simeq$5.97 Å ($\simeq$304 km s$^{-1}$) apart, a resolving power of $\lambda/\Delta\lambda \gtrsim 10^3$ suffices (for Ca II H&K, whose members are $\simeq$34.8 Å apart, $\lambda/\Delta\lambda \gtrsim 10^2$) – roughly two orders of magnitude below the $\lambda/\Delta\lambda \sim c/b$ required to resolve the velocity structure. Secondly, the inferred column density converts directly into a visual extinction through the Milky Way relation $\log N(\text{Na I}) = 13.18 + 1.125 \log A_V$ (P13): The observed range $\log N(\text{Na I}) \simeq 12$–14 corresponds to $A_V \simeq 0.1$–5 mag, so two EW measurements from a medium-resolution spectrum yield an extinction estimate over the range relevant for most Galactic and extragalactic applications. The relation carries, however, a real sightline-to-sightline scatter: The 1$\sigma$ dispersion of 0.26 dex in $\log N(\text{Na I})$ about the fit limits the precision of an individual extinction estimate to $\simeq$54 per cent of $A_V$ (P13). The conversion is therefore best viewed as a statistical extinction estimate – most powerful for large samples – although even a factor-of-two constraint is valuable for transients lacking any other extinction probe. We stress that the calibration enters only at this final step, through its two fitted coefficients: the inversion itself yields $N(\text{Na I})$ independently of any extinction calibration, so a revised $\log N(\text{Na I})$–$A_V$ relation can be substituted directly (see Section 4).

**Table 1.** Benchmark values of the inversion functions for the idealized doublet with $\mu = 2$ and $\lambda_1 = \lambda_2$ (so $r = R$ and $R_{\rm thin} = 2$): the strict-bound correction $\tilde{\kappa}_{\rm box}$, the exact Gaussian inversion ($\tilde{\kappa}$, $\tilde{\kappa}'$, $\tau_{0,2}$), and its second-order counterparts.

| $R$ | $\tilde{\kappa}_{\rm box}$ | $\tilde{\kappa}$ | $\tilde{\kappa}_{\rm 2nd}$ | $\tilde{\kappa}'$ | $\tilde{\kappa}'_{\rm 2nd}$ | $\tau_{0,2}$ | $\tau_{0,2}^{\rm 2nd}$ |
|---|---|---|---|---|---|---|---|
| 1.9 | 1.1091 | 1.1097 | 1.1053 | 2.0657 | 2.3149 | 0.3031 | 0.2694 |
| 1.7 | 1.3987 | 1.4091 | 1.3529 | 0.7401 | 1.0345 | 1.0741 | 0.7379 |
| 1.5 | 1.8484 | 1.9076 | 1.6667 | 0.4773 | 0.8311 | 2.2551 | 1.1314 |
| 1.3 | 2.6461 | 2.9889 | 2.0769 | 0.3610 | 0.7990 | 4.6710 | 1.4666 |
| 1.2 | 3.3530 | 4.4764 | 2.3333 | 0.3157 | 0.8145 | 8.0002 | 1.6162 |
| 1.1 | 4.6517 | 16.2270 | 2.6364 | 0.2469 | 0.8472 | 37.0727 | 1.7556 |

The remainder of this section presents the practical application of the formalism: the inversion functions in practice (Section 3.1) and a validation of the method against velocity-resolved measurements of Galactic sightlines (Section 3.2) and SNe Ia (Section 3.3).

### 3.1 The inversion in practice

The right-hand axis of panel (a) of Fig. 1 presents the inversion in the form used directly by an observer: the column density per unit equivalent width for the Na ID doublet, $N/\mathrm{EW}_{\rm D_2} = \kappa(R) = \tilde{\kappa}(R)/A_{\rm D_2}$, for the four estimators. In the optically thin limit, the coefficient is $1/A_{\rm D_2} \simeq 5.1 \times 10^{12}\,\mathrm{cm}^{-2}\,\text{Å}^{-1}$, and it rises steeply with saturation. In practice, an observer multiplies the measured $\mathrm{EW}_{\rm D_2}$ by the value of $\kappa$ read off at the observed doublet ratio – the strict-bound curve $\tilde{\kappa}_{\rm box}/A_{\rm D_2}$ for the recommended $N_{\rm min}$, or the exact-Gaussian curve for the model-based $N_{\tilde{\kappa}}$ (which, through panel b, also yields the Doppler parameter).

The optically thin approximation reproduces the exact Gaussian solution and $N_{\rm min}$ only in the limit $R \to 2$. The second-order estimator provides an excellent approximation throughout the optically thin and mildly saturated regimes, deviating significantly from the exact Gaussian solution only for strongly saturated absorption ($R \to 1$), where $\tilde{\kappa}(R)$ rises steeply and the inferred column density becomes increasingly sensitive to the measured doublet ratio (Appendix A). The strict-bound correction $\tilde{\kappa}_{\rm box}$ tracks the exact Gaussian one closely throughout the usable range, so the same reading applies to the recommended $N_{\rm min}$ – with the differences that it remains a valid lower bound at all $R$, and that its logarithmic growth makes it markedly less sensitive to the measured doublet ratio in the saturated regime (Appendix A).

### 3.2 Comparison with Galactic sightlines

To demonstrate the applicability of the proposed estimators, we first test them against high-resolution measurements of Galactic Na I absorption. Very high-resolution observations have shown that the Galactic Na I velocity structure is composed of narrow, closely spaced components. In particular, the $\lambda/\Delta\lambda \gtrsim 10^5$ observations of D. E. Welty et al. (1994) and I. A. Crawford (2001) resolve components with typical Doppler parameters of order $b \sim 1$ km s$^{-1}$. The high-resolution study of D. E. Welty & L. M. Hobbs (2001) similarly demonstrates that increasing spectral resolution reveals additional narrow component structure that can remain blended at lower resolution.

We therefore test explicitly whether the presence of such narrow components affects the column densities inferred from the integrated doublet equivalent widths. For this purpose, we use the Galactic Na I measurements of K. R. Sembach et al. (1993), which provide both integrated EWs for the two members of the Na ID doublet and column densities obtained from profile fitting. Their spectra were obtained with the 1.4-m Coudé Auxiliary Telescope and the Coudé Echelle Spectrograph at ESO, La Silla, at a resolving power of $\lambda/\Delta\lambda \simeq 60\,000$ (instrumental FWHM of 4.4 km s$^{-1}$), and the Na I absorption profiles were decomposed into individual velocity components. We compare the total Na I column densities obtained from these profile fits with the four estimators inferred in this work using only the integrated EWs of the Na ID doublet. Of the 57 stars in their survey, 52 have Na I measurements; excluding the two Magellanic Cloud sightlines (HD 5980 and HD 36402) leaves the 50 Galactic sightlines analysed here. These are the same 50 K. R. Sembach et al. (1993) sightlines whose profile-fitting column densities enter the fit of the $\log N(\text{Na I})$–$A_V$ relation adopted in Section 3 (equation 4 of P13), so this comparison directly probes the data underlying that relation. As the reference values, we adopt their profile-fitting totals, in analogy with the velocity-resolved profile-fitting measurements used for the SN Ia comparison in Section 3.3; their adopted apparent-optical-depth column densities are consistent with these, exceeding them only for a few strongly saturated sightlines, for which our estimators are lower bounds in any case.

We use the integrated Na ID EWs and their uncertainties reported by K. R. Sembach et al. (1993), without remeasuring the spectra. The EW uncertainties are propagated to the inferred column densities using Monte Carlo sampling ($2 \times 10^4$ Gaussian draws of the two EWs per sightline): The quoted uncertainties are the 16th–84th percentiles of the resulting distributions, for all estimators uniformly, and the quoted $\sigma_R$ is half the 16th–84th range of the $R$ draws. Draws with unphysical doublet ratios (below the saturated limit) are discarded for $N_{\tilde{\kappa}}$ and $N_{\rm min}$, whose inversions are undefined there ($N_{\rm 2nd}$ and $\sigma_R$ are computed on all draws), and draws with $R \geq R_{\rm thin}$ are assigned the optically thin correction – for every estimator, including $N_{\rm 2nd}$, so the hierarchy of equation (2) holds draw by draw. Draws requiring $\tau_{0,2} > 10^7$ are discarded for $N_{\tilde{\kappa}}$ as outside the tabulated inversion; the upper percentiles of $N_{\tilde{\kappa}}$ for strongly saturated systems are sensitive to this truncation and are quoted as indicative only (in that regime $N_{\tilde{\kappa}}$ is a lower bound only for velocity-separated components; $N_{\rm min}$ is the certified bound in general). Analytical estimates of the propagated uncertainties, useful for planning observations, are given in Appendix A.

Fig. 2 shows this comparison as a function of the observed doublet ratio. To examine explicitly whether narrow absorption components affect the comparison, we divide the sightlines according to the Doppler parameters of the fitted components. The right-hand panels show systems containing at least one component with $b_i > 3\,{\rm km\,s^{-1}}$, while the left-hand panels show the four systems for which all components satisfy $b_i \leq 3\,{\rm km\,s^{-1}}$. The upper panels compare the profile-fitting column densities of K. R. Sembach et al. (1993) with the estimators from this work, while the lower panels show their logarithmic difference. Horizontal uncertainties in $R$ are omitted for clarity.

Over the full sample, the recommended estimator, $N_{\rm min}$, closely tracks the published values: The median offset is $-0.03$ dex with an rms scatter of 0.16 dex for $R > 1.4$ ($n = 18$), and $-0.12$ dex with an rms of 0.14 dex for $1.2 < R \leq 1.4$ ($n = 27$). Only five systems have $R \leq 1.2$, with $R_{\rm min} \simeq 1.15$, and $N_{\rm min}$ lies a median 0.11 dex below the published values for these systems. Over the informative range $R > 1.2$ ($n = 45$), the residual distributions of $N_{\tilde{\kappa}}$ and $N_{\rm min}$ have medians of $-0.06$ and $-0.10$ dex, with rms scatters of 0.14 and 0.15 dex, respectively, whereas $N_{\rm 2nd}$ and $N^{\rm thin}$ are increasingly biased low, as expected from the hierarchy of equation (2). Importantly, the four sightlines for which all fitted components have $b_i \leq 3\,{\rm km\,s^{-1}}$ do not show a qualitatively different behaviour from the remainder of the sample: $N_{\rm min}$ remains close to and below the profile-fitting column densities, with residuals of approximately $\simeq -0.08$ to $-0.25$ dex. Thus, although the number of such narrow-component sightlines is small, we find no evidence within the K. R. Sembach et al. (1993) sample for an additional systematic discrepancy associated with $b_i \leq 3\,{\rm km\,s^{-1}}$.

### 3.3 Comparison with Type Ia supernova observations

We also apply the proposed estimators to the Na I D measurements of SNe Ia presented by P13. For each object, we measured the Na I $D_2$ and $D_1$ EWs,[2] and compute the optically thin, closed-form second-order, and exact Gaussian column-density estimates, together with the strict bound $N_{\rm min}$, and compare the results. The spectra were obtained from WISeREP (O. Yaron & A. Gal-Yam 2012), the ESO Science Archive (M. Romaniello & The ESO Science Archive Operations and Development Teams 2022.), and from private communications with M. Phillips. The spectrograph used for each observation is listed in Table 2; the resolving powers span $\lambda/\Delta\lambda \simeq 1$–$4.5 \times 10^4$ (Magellan/MIKE $\simeq 3.5 \times 10^4$, R. Bernstein et al. 2003; Keck/HIRES $\simeq 4.5 \times 10^4$, S. S. Vogt et al. 1994; VLT/UVES $\simeq 4 \times 10^4$, S. D'Odorico et al. 2000; HET/HRS $\simeq 3 \times 10^4$, R. G. Tull 1998; and VLT/X-shooter $\simeq 1 \times 10^4$, J. Vernet et al. 2011). Our analysis is restricted to the spectra in hand: Several SNe from the P13 sample are not included, as their spectra neither are publicly available nor were provided to us. For the comparison presented below, we restrict the sample to absorption systems with significant detections in both doublet members, requiring ${\rm EW_{D_1}} > 3\,\sigma_{\rm EW_{D_1}}$ and ${\rm EW_{D_2}} > 3\,\sigma_{\rm EW_{D_2}}$.

The uncertainties in the measured EWs are propagated to the inferred column densities using the same Monte Carlo procedure described in Section 3.2. The measurements and inferred column densities are summarized in Table 2. A machine-readable version of the table is provided as supplementary material.

Fig. 3 compares the column densities obtained with the four estimators to those reported by P13. Horizontal uncertainties in $R$ are omitted for clarity. We describe first the recommended estimator. Over the full sample of 36 systems, the median offset of $\log N_{\rm min}$ from the published values is $-0.08$ dex with an rms scatter of 0.13 dex for $R > 1.4$ ($n = 17$), and $-0.14$ dex with an rms of 0.18 dex for $1.2 < R < 1.4$ ($n = 11$). For the eight strongly saturated systems with $R < 1.2$, the strict bound lies a median 0.6 dex below the published values, as expected for a lower limit. The excursions of $N_{\rm min}$ above published values are all statistically insignificant or marginal: The largest is $+0.16$ dex at $0.7\sigma$ (SN 2009le), and the most significant is $+0.07$ dex at $2.3\sigma$ (SN 2011ek). The exact Gaussian inversion closely reproduces the published values wherever the doublet ratio is informative – median offsets of $-0.08$ and $-0.11$ dex with rms scatters of 0.13 and 0.19 dex in the two ranges above, nearly identical to $N_{\rm min}$ since $\eta \leq 1.06$ for $R > 1.4$ – while for $R < 1.2$ it scatters around them with an rms of 0.7 dex. Overall, 83 per cent of the $N_{\tilde{\kappa}}$ values agree with the published ones within twice the combined uncertainties. The closed-form second-order estimator consistently yields lower column densities, whereas the optically thin approximation yields the lowest values. The marginal histogram in the bottom-right panel of Fig. 3 summarizes these residuals for the systems with informative doublet ratios ($R > 1.2$, $n = 28$): the distributions of $N_{\tilde{\kappa}}$ and $N_{\rm min}$ cluster around zero (medians of $-0.09$ and $-0.12$ dex, with an rms of 0.16 dex), whereas those of $N_{\rm 2nd}$ and $N^{\rm thin}$ are increasingly biased low, as expected for the lower bounds of equation (2). We emphasize that the profile-fitting results of P13 are based on high-resolution spectroscopy that resolves the velocity structure, whereas our estimates use only the two integrated EWs.

The internal consistency diagnostic proposed in Section 2.3 is directly visible in the data: for $R \gtrsim 1.4$, $N_{\tilde{\kappa}}$ and $N_{\rm 2nd}$ agree to better than $\sim$0.1 dex, and both reproduce the published values; towards stronger saturation the two estimators diverge, with $\log(N_{\tilde{\kappa}}/N_{\rm 2nd})$ reaching $\sim$0.8 dex at $R \simeq 1.1$, correctly flagging the objects for which the saturation correction, and hence the systematic uncertainty, is large. The ratio $N_{\tilde{\kappa}}/N_{\rm 2nd}$ thus provides a per-object accuracy estimate that requires no information beyond the measured EWs.

[2]Measurement uncertainties include both statistical and systematic contributions. Further details of the measurement procedure and uncertainty estimation will be presented in Badash & Kushnir (in preparation).

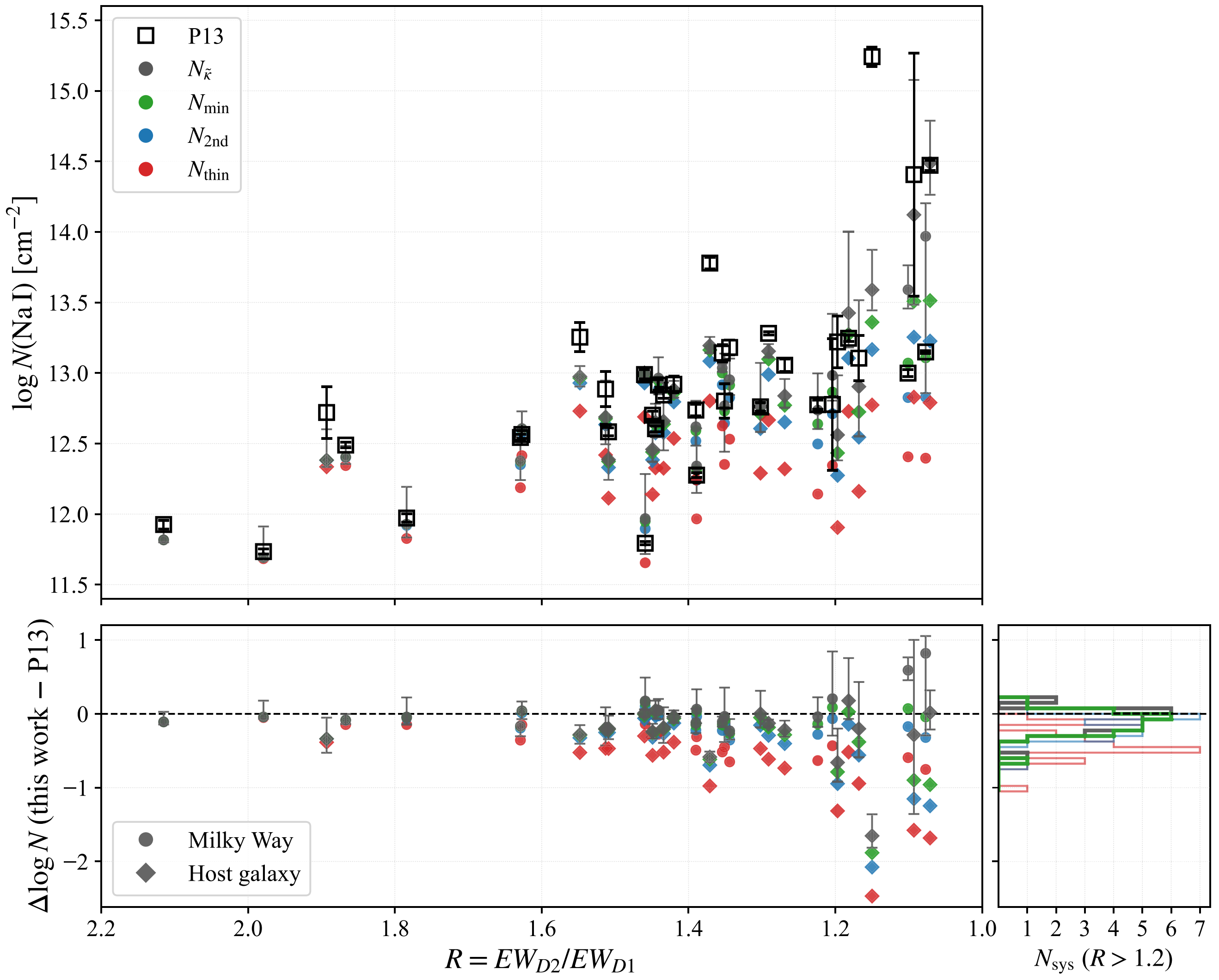

**Figure 3.** Comparison of Na I column densities inferred for the P13 sample using the four estimators discussed in this work. Top: Column densities derived from the exact Gaussian inversion $N_{\tilde{\kappa}}$ (grey), the recommended strict bound $N_{\min}$ (green), the closed-form second-order estimator $N_{\rm 2nd}$ (blue), and the optically thin approximation $N^{\rm thin}$ (red), compared with the values reported by P13 (open black squares); the estimator colours follow the curves of Fig. 1. Bottom: Difference between each estimate and the published values as a function of the observed doublet ratio $R$; the error bars here combine our uncertainties with the published ones in quadrature. Circles denote Milky Way absorption systems, while diamonds denote the SNe Ia host-galaxy absorption systems. All estimators agree in the optically thin regime ($R \approx 2$), while their differences increase with saturation ($R \to 1$), preserving the hierarchy of equation (2). For clarity, error bars are shown only for $N_{\tilde{\kappa}}$ and for the published values; they represent the 16th–84th percentile confidence intervals obtained by Monte Carlo propagation of the measured equivalent-width uncertainties. Bottom right: Histograms of the residuals for the systems with informative doublet ratios ($R > 1.2$; $n = 28$), in the same colours: $N_{\tilde{\kappa}}$ and $N_{\min}$ (thick) cluster around zero (median $-0.09$ and $-0.12$ dex, rms 0.16 dex), whereas $N_{\rm 2nd}$ and $N^{\rm thin}$ (thin) are increasingly biased low, as expected for the lower bounds of equation (2).

### 3.4 Summary of the algorithm and an explicit extinction formula

For an observer, the entire procedure reduces to the following steps.

(i) Measure the equivalent widths of the two doublet members, $\mathrm{EW}_{\mathrm{D}_2}$ and $\mathrm{EW}_{\mathrm{D}_1}$, and their uncertainties. Both members should be detected significantly. The only resolution requirement is that the two members be measured separately ($\lambda/\Delta\lambda \gtrsim 10^3$ for Na ID ); for extragalactic sources, the host and Milky Way systems must be separated by the source redshift.

(ii) Form the doublet ratio $R = \mathrm{EW}_{\mathrm{D}_2}/\mathrm{EW}_{\mathrm{D}_1}$ and locate it in the usable range: $R \geq 2$ indicates optically thin absorption, for which $N = N^{\rm thin} = 5.1 \times 10^{12}\,\mathrm{cm}^{-2}\,(\mathrm{EW}_{\mathrm{D}_2}/\text{Å})$; $R \lesssim 1.1$ indicates saturation too strong for a useful estimate ($N_{\min}$ below remains a valid lower bound); values outside the physical range $1 < R < 2$ signal measurement uncertainties.

(iii) Compute the recommended column density – the strict bound $N_{\min} = N^{\rm thin}\,\tilde{\kappa}_{\rm box}(R)$, with the closed form $\tilde{\kappa}_{\rm box}(R) = 2\ln\left[1/(R-1)\right]/\left[1-(R-1)^2\right]$ for Na ID (Section 3) – and, if the single-Gaussian model is adopted, the exact inversion $N_{\tilde{\kappa}} = N^{\rm thin}\,\tilde{\kappa}(R)$ and the Doppler parameter $b$ from Fig. 1.

(iv) Assess the saturation strength through the ratio $N_{\tilde{\kappa}}/N_{\rm 2nd}$, with $N_{\rm 2nd} = N^{\rm thin}\,(4/R-1)$ (equation 37): A ratio near unity certifies a mildly saturated system for which $N_{\min}$ is accurate, while

**Table 2.** Measured Na I D equivalent widths and inferred column densities for the P13 sample used in Fig. 3. For each object we list the measured equivalent widths of the $D_2$ and $D_1$ transitions, the column density reported by P13, and the column densities inferred using the recommended strict lower bound $N_{\rm min}$ (Section 2.3), the exact Gaussian inversion, the closed-form second-order estimator, and the optically thin approximation, in the order of the columns. All derived quantities ($R$ and the four column densities) are computed from the measured EWs with a single Monte Carlo pipeline ($2 \times 10^4$ Gaussian draws of the two EWs per system); the quoted uncertainties are the 16th–84th percentiles, and $\sigma_R$ is half the 16th–84th range of the $R$ draws, with draws with unphysical doublet ratios discarded (for $N_{\tilde{\kappa}}$ and $N_{\rm min}$, whose inversions are undefined there; $N_{\rm 2nd}$ and $\sigma_R$ use all draws) and draws with $R \geq R_{\rm thin}$ assigned the optically thin correction for all estimators uniformly (in particular $\tilde{\kappa}_{\rm 2nd} = 1$ there, so $N^{\rm thin} \leq N_{\rm 2nd}$ holds draw by draw). For $N_{\tilde{\kappa}}$, draws requiring $\tau_{0,2} > 10^7$ are discarded as outside the tabulated inversion; for the strongly saturated systems ($R < 1.2$) the quoted upper percentile of $N_{\tilde{\kappa}}$ is sensitive to this truncation and should be regarded as indicative only – in that regime $N_{\tilde{\kappa}}$ is a lower bound only for velocity-separated components, and $N_{\rm min}$ is the certified bound in general (Section 2.4). Milky Way and host-galaxy absorption systems are listed separately.

| SN name | Telescope/instrument | $EW_{D2}$ (Å) | $EW_{D1}$ (Å) | $R$ | $\log N_{\rm Na\,I}$ (P13) (cm$^{-2}$) | $\log N_{\rm min}$ (cm$^{-2}$) | $\log N_{\tilde{\kappa}}$ (cm$^{-2}$) | $\log N_{\rm 2nd}$ (cm$^{-2}$) | $\log N_{\rm thin}$ (cm$^{-2}$) |
|---|---|---|---|---|---|---|---|---|---|
| | | | | *Milky Way* | | | | | |
| SN 2003gd | Magellan/MIKE | $0.273 \pm 0.013$ | $0.223 \pm 0.013$ | $1.224 \pm 0.091$ | $12.775 \pm 0.034$ | $12.639^{+0.095}_{-0.082}$ | $12.737^{+0.260}_{-0.134}$ | $12.497 \pm 0.037$ | $12.142 \pm 0.021$ |
| SN 2006ca | Keck/HIRES | $0.668 \pm 0.035$ | $0.497 \pm 0.036$ | $1.344 \pm 0.121$ | $13.181 \pm 0.052$ | $12.914^{+0.096}_{-0.086}$ | $12.952^{+0.150}_{-0.108}$ | $12.826 \pm 0.049$ | $12.531 \pm 0.023$ |
| SN 2006eu | Keck/HIRES | $0.853 \pm 0.055$ | $0.592 \pm 0.055$ | $1.441 \pm 0.163$ | $12.914 \pm 0.039$ | $12.944^{+0.111}_{-0.102}$ | $12.964^{+0.147}_{-0.116}$ | $12.886 \pm 0.065$ | $12.637 \pm 0.028$ |
| SN 2007kk | Keck/HIRES | $0.443 \pm 0.078$ | $0.328 \pm 0.082$ | $1.351 \pm 0.428$ | $12.801 \pm 0.122$ | $12.730^{+0.207}_{-0.287}$ | $12.767^{+0.370}_{-0.325}$ | $12.645 \pm 0.165$ | $12.352 \pm 0.077$ |
| SN 2007sr | Keck/HIRES | $0.095 \pm 0.009$ | $0.048 \pm 0.012$ | $1.979 \pm 0.557$ | $11.734 \pm 0.018$ | $11.692^{+0.211}_{-0.019}$ | $11.692^{+0.220}_{-0.019}$ | $11.692 \pm 0.098$ | $11.684 \pm 0.041$ |
| SN 2008C | Keck/HIRES | $0.436 \pm 0.025$ | $0.362 \pm 0.026$ | $1.204 \pm 0.111$ | $12.777 \pm 0.467$ | $12.865^{+0.124}_{-0.104}$ | $12.983^{+0.436}_{-0.179}$ | $12.711 \pm 0.046$ | $12.345 \pm 0.025$ |
| SN 2008fp | Magellan/MIKE | $0.830 \pm 0.010$ | $0.613 \pm 0.011$ | $1.354 \pm 0.029$ | $13.141 \pm 0.061$ | $13.000^{+0.022}_{-0.022}$ | $13.036^{+0.030}_{-0.028}$ | $12.916 \pm 0.012$ | $12.625 \pm 0.005$ |
| SN 2008hv | Magellan/MIKE | $0.182 \pm 0.019$ | $0.131 \pm 0.019$ | $1.389 \pm 0.250$ | $12.276 \pm 0.016$ | $12.312^{+0.172}_{-0.166}$ | $12.341^{+0.267}_{-0.191}$ | $12.240 \pm 0.100$ | $11.966 \pm 0.045$ |
| SN 2008ia | Magellan/MIKE | $0.490 \pm 0.065$ | $0.455 \pm 0.053$ | $1.077 \pm 0.191$ | $13.149 \pm 0.010$ | $13.107^{+0.066}_{-0.282}$ | $13.969^{+0.234}_{-1.112}$ | $12.829 \pm 0.072$ | $12.396 \pm 0.057$ |
| SN 2009ds | Magellan/MIKE | $0.435 \pm 0.019$ | $0.233 \pm 0.020$ | $1.867 \pm 0.180$ | $12.489 \pm 0.020$ | $12.405^{+0.077}_{-0.050}$ | $12.405^{+0.079}_{-0.050}$ | $12.402 \pm 0.056$ | $12.344 \pm 0.019$ |
| SN 2009ev | Magellan/MIKE | $0.346 \pm 0.025$ | $0.249 \pm 0.025$ | $1.390 \pm 0.171$ | $12.737 \pm 0.040$ | $12.591^{+0.125}_{-0.114}$ | $12.619^{+0.184}_{-0.134}$ | $12.518 \pm 0.069$ | $12.245 \pm 0.032$ |
| SN 2009iw | Magellan/MIKE | $0.303 \pm 0.027$ | $0.186 \pm 0.027$ | $1.629 \pm 0.281$ | $12.543 \pm 0.021$ | $12.373^{+0.147}_{-0.132}$ | $12.378^{+0.170}_{-0.137}$ | $12.350 \pm 0.104$ | $12.187 \pm 0.039$ |
| SN 2009le | Magellan/MIKE | $0.089 \pm 0.014$ | $0.061 \pm 0.015$ | $1.459 \pm 0.451$ | $11.793 \pm 0.011$ | $11.950^{+0.223}_{-0.233}$ | $11.968^{+0.316}_{-0.251}$ | $11.896 \pm 0.159$ | $11.655 \pm 0.068$ |
| SN 2009mz | Magellan/MIKE | $0.132 \pm 0.019$ | $0.074 \pm 0.019$ | $1.784 \pm 0.556$ | $11.972 \pm 0.030$ | $11.928^{+0.237}_{-0.093}$ | $11.929^{+0.264}_{-0.095}$ | $11.920 \pm 0.128$ | $11.827 \pm 0.063$ |
| SN 2009nr | Magellan/MIKE | $0.129 \pm 0.012$ | $0.061 \pm 0.013$ | $2.115 \pm 0.510$ | $11.926 \pm 0.031$ | $11.817^{+0.130}_{-0.017}$ | $11.817^{+0.133}_{-0.017}$ | $11.817 \pm 0.067$ | $11.817 \pm 0.040$ |
| SN 2010ev | Magellan/MIKE | $0.511 \pm 0.036$ | $0.314 \pm 0.033$ | $1.627 \pm 0.205$ | $12.564 \pm 0.028$ | $12.601^{+0.108}_{-0.106}$ | $12.606^{+0.122}_{-0.110}$ | $12.578 \pm 0.080$ | $12.414 \pm 0.030$ |
| SN 2011ek | VLT/X-shooter | $0.501 \pm 0.004$ | $0.455 \pm 0.004$ | $1.101 \pm 0.013$ | $12.999 \pm 0.024$ | $13.070^{+0.021}_{-0.020}$ | $13.590^{+0.173}_{-0.133}$ | $12.826 \pm 0.005$ | $12.406 \pm 0.003$ |
| | | | | *Host galaxy* | | | | | |
| SN 2001el | VLT/UVES | $0.384 \pm 0.033$ | $0.295 \pm 0.033$ | $1.302 \pm 0.185$ | $12.760 \pm 0.030$ | $12.710^{+0.152}_{-0.141}$ | $12.762^{+0.309}_{-0.180}$ | $12.606 \pm 0.075$ | $12.290 \pm 0.037$ |
| SN 2002bo | HET/HRS | $1.329 \pm 0.077$ | $1.216 \pm 0.076$ | $1.093 \pm 0.093$ | $14.406 \pm 0.862$ | $13.508^{+0.117}_{-0.139}$ | $14.121^{+0.956}_{-0.637}$ | $13.254 \pm 0.038$ | $12.829 \pm 0.025$ |
| SN 2002ha | HET/HRS | $0.516 \pm 0.062$ | $0.341 \pm 0.062$ | $1.513 \pm 0.339$ | $12.886 \pm 0.125$ | $12.676^{+0.196}_{-0.181}$ | $12.688^{+0.258}_{-0.193}$ | $12.634 \pm 0.128$ | $12.419 \pm 0.052$ |
| SN 2002jg | HET/HRS | $1.052 \pm 0.077$ | $0.890 \pm 0.076$ | $1.182 \pm 0.134$ | $13.246 \pm 0.021$ | $13.274^{+0.138}_{-0.134}$ | $13.425^{+0.576}_{-0.245}$ | $13.105 \pm 0.054$ | $12.728 \pm 0.032$ |
| SN 2006X | HET/HRS | $1.248 \pm 0.034$ | $0.910 \pm 0.033$ | $1.371 \pm 0.062$ | $13.779 \pm 0.041$ | $13.163^{+0.045}_{-0.043}$ | $13.194^{+0.061}_{-0.053}$ | $13.084 \pm 0.025$ | $12.802 \pm 0.012$ |
| SN 2006cm | Keck/HIRES | $1.166 \pm 0.028$ | $1.014 \pm 0.029$ | $1.150 \pm 0.043$ | $15.242 \pm 0.069$ | $13.360^{+0.058}_{-0.050}$ | $13.589^{+0.284}_{-0.145}$ | $13.166 \pm 0.018$ | $12.773 \pm 0.010$ |
| SN 2007fb | Keck/HIRES | $0.416 \pm 0.046$ | $0.290 \pm 0.048$ | $1.434 \pm 0.294$ | $12.844 \pm 0.024$ | $12.637^{+0.187}_{-0.186}$ | $12.658^{+0.274}_{-0.206}$ | $12.577 \pm 0.117$ | $12.325 \pm 0.048$ |
| SN 2007fs | Keck/HIRES | $0.255 \pm 0.020$ | $0.169 \pm 0.021$ | $1.509 \pm 0.224$ | $12.583 \pm 0.029$ | $12.373^{+0.137}_{-0.132}$ | $12.386^{+0.171}_{-0.143}$ | $12.330 \pm 0.090$ | $12.113 \pm 0.034$ |
| SN 2007le | Keck/HIRES | $0.915 \pm 0.017$ | $0.709 \pm 0.016$ | $1.291 \pm 0.037$ | $13.281 \pm 0.012$ | $13.098^{+0.031}_{-0.030}$ | $13.154^{+0.050}_{-0.043}$ | $12.989 \pm 0.015$ | $12.667 \pm 0.008$ |

**Table 2** – *continued*

| SN name | Telescope/instrument | $EW_{D2}$ (Å) | $EW_{D1}$ (Å) | $R$ | $\log N_{\rm Na\,I}$ (P13) ($cm^{-2}$) | $\log N_{\rm min}$ ($cm^{-2}$) | $\log N_{\tilde{\kappa}}$ ($cm^{-2}$) | $\log N_{\rm 2nd}$ ($cm^{-2}$) | $\log N_{\rm thin}$ ($cm^{-2}$) |
|---|---|---|---|---|---|---|---|---|---|
| | | | | | *Milky Way* | | | | |
| SN 2007sr | Keck/HIRES | $0.158 \pm 0.008$ | $0.132 \pm 0.009$ | $1.197 \pm 0.102$ | $13.220 \pm 0.182$ | $12.433^{+0.115}_{-0.101}$ | $12.561^{+0.422}_{-0.180}$ | $12.274 \pm 0.043$ | $11.905 \pm 0.022$ |
| SN 2008C | Keck/HIRES | $0.426 \pm 0.051$ | $0.225 \pm 0.052$ | $1.893 \pm 0.519$ | $12.720 \pm 0.184$ | $12.383^{+0.204}_{-0.049}$ | $12.383^{+0.218}_{-0.049}$ | $12.381 \pm 0.103$ | $12.335 \pm 0.052$ |
| SN 2008ec | Keck/HIRES | $0.411 \pm 0.014$ | $0.324 \pm 0.014$ | $1.269 \pm 0.069$ | $13.055 \pm 0.044$ | $12.771^{+0.063}_{-0.056}$ | $12.838^{+0.119}_{-0.083}$ | $12.652 \pm 0.028$ | $12.320 \pm 0.015$ |
| SN 2008fp | Magellan/MIKE | $1.214 \pm 0.008$ | $1.133 \pm 0.008$ | $1.071 \pm 0.010$ | $14.472 \pm 0.036$ | $13.513^{+0.022}_{-0.020}$ | $14.489^{+0.299}_{-0.226}$ | $13.226 \pm 0.004$ | $12.790 \pm 0.003$ |
| SN 2009ds | Magellan/MIKE | $0.676 \pm 0.020$ | $0.476 \pm 0.020$ | $1.420 \pm 0.072$ | $12.920 \pm 0.058$ | $12.859^{+0.048}_{-0.047}$ | $12.882^{+0.062}_{-0.056}$ | $12.795 \pm 0.029$ | $12.536 \pm 0.013$ |
| SN 2009ig | Magellan/MIKE | $0.285 \pm 0.028$ | $0.244 \pm 0.029$ | $1.168 \pm 0.184$ | $13.105 \pm 0.161$ | $12.724^{+0.143}_{-0.195}$ | $12.903^{+0.613}_{-0.350}$ | $12.545 \pm 0.075$ | $12.161 \pm 0.043$ |
| SN 2009le | Magellan/MIKE | $1.056 \pm 0.044$ | $0.682 \pm 0.042$ | $1.548 \pm 0.115$ | $13.254 \pm 0.103$ | $12.964^{+0.065}_{-0.065}$ | $12.974^{+0.075}_{-0.070}$ | $12.929 \pm 0.045$ | $12.730 \pm 0.018$ |
| SN 2010A | Magellan/MIKE | $0.419 \pm 0.025$ | $0.290 \pm 0.025$ | $1.445 \pm 0.150$ | $12.605 \pm 0.021$ | $12.633^{+0.099}_{-0.094}$ | $12.652^{+0.130}_{-0.107}$ | $12.575 \pm 0.060$ | $12.328 \pm 0.026$ |
| SN 2010ev | Magellan/MIKE | $0.271 \pm 0.014$ | $0.187 \pm 0.013$ | $1.449 \pm 0.125$ | $12.701 \pm 0.029$ | $12.440^{+0.081}_{-0.077}$ | $12.459^{+0.104}_{-0.087}$ | $12.384 \pm 0.049$ | $12.139 \pm 0.022$ |
| SN 2012cg | VLT/X-shooter | $0.961 \pm 0.019$ | $0.658 \pm 0.019$ | $1.460 \pm 0.051$ | $12.989 \pm 0.035$ | $12.982^{+0.032}_{-0.032}$ | $13.000^{+0.038}_{-0.037}$ | $12.929 \pm 0.020$ | $12.689 \pm 0.008$ |

a large ratio flags strong saturation – for $R < 1.4$, quote $N_{\min}$ as a lower limit.

(v) Propagate the EW uncertainties – by Monte Carlo through the exact mapping (Section 3.2) or analytically (Appendix A) – to obtain the confidence interval of the column density; their impact is amplified as $R \to 1$.

(vi) If an extinction estimate is desired, convert through the Milky Way relation of P13 (Section 3): The full chain of this paper collapses to the single explicit formula

$$A_V \simeq 0.38 \left[\tilde{\kappa}_{\rm box}(R)\,\frac{{\rm EW}_{\rm D_2}}{\text{Å}}\right]^{1/1.125} \text{mag},$$
$$\tilde{\kappa}_{\rm box}(R) = \frac{2\ln\left[1/(R-1)\right]}{1-(R-1)^2}, \qquad R = \frac{{\rm EW}_{\rm D_2}}{{\rm EW}_{\rm D_1}}, \tag{38}$$

which in the optically thin limit ($R \to 2$, $\tilde{\kappa}_{\rm box} \to 1$) reduces to $A_V \simeq 0.38\,({\rm EW}_{\rm D_2}/\text{Å})^{0.89}$ mag. Three caveats accompany the conversion: for $1.1 \lesssim R \lesssim 1.4$ the result must be quoted as a lower limit, which the true extinction may exceed by up to a factor $\eta(R)^{1/1.125}$, growing to $\simeq 3$ at $R = 1.1$ (Section 2.4); the calibration spans $\log N(\text{Na I}) \simeq 12$–14, i.e. $A_V \simeq 0.1$–5 mag; and even for perfectly measured EWs, the sight-line scatter of the Milky Way relation (0.26 dex in $\log N$, i.e. 0.23 dex in $\log A_V$) limits the precision of a single line of sight to $\simeq 54$ per cent of $A_V$ (Section 3).

All steps are implemented in the public reference code (Data Availability): `columns_NaID(EW2, EW1, sigma2, sigma1)` returns the four estimators, the Doppler parameter, and their 16th–84th percentile uncertainties, propagated exactly as in Section 3.2, while `AV_NaID` implements equation (38) with the exact Na ID atomic parameters and additionally returns $A_V$ with both the measurement-only and the total (including the Milky Way-relation scatter) uncertainties, warning when the estimate falls outside the calibrated range.

## 4 SUMMARY AND DISCUSSION

We have shown that the Gaussian COG inversion for a resonance-line doublet can be reformulated as a universal saturation correction depending only on the observed doublet ratio $R$: For any fixed $R$, the inferred column density is simply proportional to the measured EW of one member of the doublet (equation 17). This reformulation yields an exact one-dimensional Gaussian inversion, a simple closed-form second-order approximation, a certified optimal lower bound, $N_{\min}$, attained by a box-shaped optical-depth profile and valid for an arbitrary velocity structure, and the hierarchy of lower bounds of equation (2). We validated these estimators against Galactic Na ID measurements (Section 3.2) and Na ID measurements of SNe Ia (Section 3.3).

The one-dimensional nature of the inversion follows from the scaling symmetry identified in Section 2.1: Rescaling the column density and the Doppler parameter together, $(N, b) \to (\alpha N, \alpha b)$, leaves the line-centre optical depth, and hence the doublet ratio, unchanged, while the equivalent widths of both members of the doublet scale by the same factor $\alpha$. The inferred column density must therefore be proportional to the measured equivalent width at fixed doublet ratio. The argument requires only a line profile characterized by a single width parameter, with $\tau_0 \propto N/b$, and is not restricted to the Gaussian case.

In practice, we recommend reporting $N_{\min}$ (equation 36) as the column density: It is a certified, optimal lower limit for an arbitrary velocity structure, and for $R \gtrsim 1.4$ it agrees with the exact Gaussian inversion to better than a few per cent – so essentially no accuracy is sacrificed for its robustness where the two agree; in the strongly saturated regime, the bound is conservative by construction. The exact Gaussian inversion, $N_{\tilde{\kappa}}$, retains two distinct roles: It is the exact solution of the classical doublet-ratio problem – recovering the true column density and Doppler parameter for single-Gaussian absorption, and bounding the column density from below for velocity-separated components – and it is the only estimator that yields the Doppler parameter, for which no distribution-free analogue of $N_{\min}$ exists (velocity information is intrinsically model-dependent). Its reliability can be assessed, object by object, by comparing the exact Gaussian inversion with the closed-form second-order estimator: The ratio $N_{\tilde{\kappa}}/N_{\rm 2nd}$ measures the strength of the saturation correction, so that $N_{\tilde{\kappa}}/N_{\rm 2nd} \approx 1$ certifies a mildly saturated system for which $N_{\tilde{\kappa}}$ (and hence also $N_{\min}$) is accurate, while a large ratio flags a strongly saturated system. This diagnostic requires nothing beyond the two measured EWs, and we applied it both analytically (Sections 2.4 and 3.1) and to the Galactic sight-line sample (Section 3.2) and the observed SNe Ia sample (Section 3.3). Recovering the true column density in the strongly saturated regime, however, requires additional information, such as velocity-resolved spectroscopy or independent unsaturated transitions.

Our analytical treatment complements both the numerical study of E. B. Jenkins (1986), who showed that ensembles of unresolved Gaussian components closely reproduce the classical Gaussian COG, and methods based on velocity-resolved spectroscopy, such as profile fitting and the AOD corrections of B. D. Savage & K. R. Sembach (1991) and E. B. Jenkins (1996). Whereas the latter exploit the full absorption profile, our method is intended for spectra in which only the integrated EWs of the doublet are available. The practical consequence is the resolution requirement quantified in Section 3: Separating the Na ID doublet members requires $\lambda/\Delta\lambda \gtrsim 10^3$, roughly two orders of magnitude below the $\lambda/\Delta\lambda \sim c/b \approx 3\times10^4$ to $3\times10^5$ needed to resolve the velocity structure of cold interstellar gas. In Sections 3.2 and 3.3, we showed that essentially no accuracy is sacrificed outside the strongly saturated regime: The proposed estimators reproduce the high-resolution profile-fitting results of K. R. Sembach et al. (1993) and P13 from the integrated EWs alone.

This resolution advantage makes the formalism particularly well-suited for upcoming medium-resolution spectroscopic surveys. Instruments such as Son-Of-XShooter (SOXS) (P. Schipani et al. 2018; K. K. R. Santhakumari et al. 2024; Ben-Ami et al. 2026 on the ESO–NTT and the planned HighSpec spectrograph on the MAST array (Y. Sofer Rimalt et al. 2024) are well suited for building large, homogeneous samples of Na ID measurements. Indeed, Badash et al. (in preparation) demonstrate that medium-resolution SOXS spectroscopy routinely resolves the Na ID doublet, while the underlying velocity structure often remains unresolved, making such observations an ideal application of the present formalism. Combined with the $\log N(\text{Na I})$–$A_V$ relation of P13 (Section 3), the method turns two EW measurements into an extinction estimate through the explicit formula of equation (38), providing a practical extinction probe for large samples of Galactic and extragalactic objects – particularly transients – for which no other method is available.

Whether the Galactic $\log N(\text{Na I})$–$A_V$ relation applies unchanged in SN Ia host galaxies remains an open question: P13 found a substantially larger dispersion when comparing host-galaxy $N(\text{Na I})$ with the reddening inferred from SN colours, with a quarter of their sample showing anomalously strong Na ID absorption. Proposed explanations include outflowing circumstellar gas from the progenitor system, which P13 suggested is responsible for at least some of the anomalously strong systems (found preferentially among SNe with blueshifted Na ID profiles), unrecognized saturation in the Na I measurements (D. E. Welty et al. 2014), and systematic uncertainties in the colour-based extinction estimates (Badash & Kushnir, in preparation). D. E. Welty et al. (2014) suggested that unrecognized saturation effects in the Galactic Na I data compiled by B. Y. Welsh et al. (2010) biased the slopes of the fiducial relations of P13, and that the true slope of the $\log N(\mathrm{Na\,I})$–$\log A_V$ relation is closer to the nearly quadratic one exhibited by $N(\mathrm{K\,I})$ (D. E. Welty & L. M. Hobbs 2001). We note, however, that the relation used here (equation 4 of P13) was not derived from the B. Y. Welsh et al. (2010) compilation: It was fit to high-resolution profile-fitting measurements in which the saturated Na ID components were constrained using the weak K I lines, combined with the K. R. Sembach et al. (1993) sample. The extinction estimates entering this relation were obtained differently for the two Galactic samples: The $A_V$ values adopted by P13 for the Galactic SN sightlines were based on the dust maps of D. J. Schlegel, D. P. Finkbeiner & M. Davis (1998), recalibrated by E. F. Schlafly & D. P. Finkbeiner (2011), whereas K. R. Sembach et al. (1993) used reddening estimates towards the individual background stars. The B. Y. Welsh et al. (2010) data enter P13 only in their comparison of the λ5780 diffuse interstellar band with $N(\mathrm{Na\,I})$. The formalism presented here is well-positioned to sharpen this comparison: It is agnostic to the adopted $N$–$A_V$ calibration, and it supplies a built-in per-object saturation diagnostic ($N_{\tilde{\kappa}}/N_{\mathrm{2nd}}$) that directly addresses the saturation concern. More importantly, the validity of the P13 calibration does not affect the main results of this paper: The inversion of Sections 2 and 3 yields $N(\mathrm{Na\,I})$ from the two EWs alone, independently of any extinction calibration, and the conversion to $A_V$ (equation 38) simply adopts the coefficients of the chosen $\log N(\mathrm{Na\,I})$–$\log A_V$ relation. Should the Galactic relation be revised – for example towards the steeper slope advocated by D. E. Welty et al. (2014) – the updated coefficients can be inserted directly, leaving the inferred column densities and the method unchanged.

More broadly, the present formalism is readily applicable to any resonance-line doublet. The separation of the inversion into a linear dependence on the measured equivalent width and a universal saturation correction that depends only on the observed doublet ratio is independent of the specific atomic species, and holds for any self-similar line profile with a single width parameter (Section 2.1, which also delineates when the Doppler-dominated Gaussian form adopted here is adequate). The formalism can therefore be applied directly to other commonly observed doublets, such as Ca II H&K, for which the required resolving power is only $\lambda/\Delta\lambda \gtrsim 10^2$, K I λλ7665, 7699, although the stronger member, λ7665, lies within the telluric $O_2$ *A* band, so sufficient spectral resolution is required in practice to separate the interstellar absorption from the telluric features, and the ultraviolet resonance doublets widely used in quasar-absorption studies (e.g. Mg II λλ2796, 2803, C IV λλ1548, 1550, and Si IV λλ1394, 1403, all with $f$-ratios close to two), providing a simple and inexpensive method for estimating column densities whenever the two members of a doublet are spectrally resolved. When several doublets of the same gas are available along a line of sight (e.g. Na ID together with K I λλ7665, 7699, or Ca II H&K), each provides an independent $(N, b)$ constraint on the same velocity field; their combination tests the single-width assumption underlying the exact inversion and, since the doublets saturate at different column densities, extends the usable range of the method and tightens the saturation correction. We leave a joint multidoublet formulation to future work.

## ACKNOWLEDGEMENTS

We thank the anonymous referee for valuable comments and suggestions that improved the clarity of the manuscript. We are grateful to Mark Phillips and his colleagues for sharing their spectra with us. We thank Mark Phillips, Avishay Gal-Yam, Boaz Katz, and Eran Ofek for their valuable discussions. We also extend our gratitude to Daniella van der Boom and Tal Wasserman for their insightful contributions. DK was supported by a research grant from The Abramson Family Center for Young Scientists, an ISF grant, the Minerva Stiftung, and the Pazi Foundation.

## DATA AVAILABILITY

A machine-readable version of Table 2 is available as supplementary material and in the repository below (`data/`). It contains the measured Na ID equivalent widths, their uncertainties, the inferred column densities from the four estimators presented in this work, and the corresponding uncertainties. A self-contained reference implementation of the inversion functions of Section 2 is available at https://github.com/DoronKushnir/doublet_inversion.

## SUPPLEMENTARY MATERIAL

Supplementary data are available at *MNRAS* online.

Please note: Oxford University Press is not responsible for the content or functionality of any supporting materials supplied by the authors. Any queries (other than missing material) should be directed to the corresponding author for the article.

## APPENDIX A: PROPAGATION OF MEASUREMENT UNCERTAINTIES

The optically thin and closed-form second-order estimators are linear in the measured EWs, so their uncertainties follow from standard linear propagation. For the exact Gaussian inversion, writing $\ln N_{\tilde{\kappa}} = \ln \mathrm{EW}_2 - \ln A_2 + \ln \tilde{\kappa}(R)$ and defining the logarithmic slope of the saturation correction,

$$\gamma(R) \equiv \frac{\mathrm{d}\ln\tilde{\kappa}}{\mathrm{d}\ln R} < 0, \tag{A1}$$

first-order propagation of independent uncertainties $\sigma_{\mathrm{EW}_1}$ and $\sigma_{\mathrm{EW}_2}$ gives

$$\sigma^2_{\ln N_{\tilde{\kappa}}} = (1+\gamma)^2\left(\frac{\sigma_{\mathrm{EW}_2}}{\mathrm{EW}_2}\right)^2 + \gamma^2\left(\frac{\sigma_{\mathrm{EW}_1}}{\mathrm{EW}_1}\right)^2. \tag{A2}$$

The factor $(1+\gamma)$ reflects a partial cancellation: Increasing $\mathrm{EW}_2$ raises the optically thin estimate but also raises $R$, which lowers the saturation correction. Since $|\gamma|$ grows steeply with saturation – for the Na ɪD doublet, $\gamma \simeq -2.1, -3.4, -6.8$, and $-31$ at $R = 1.9, 1.35, 1.2$, and 1.1, respectively – equation (A2) quantifies how the achievable column-density precision degrades as $R \to 1$, and converts a target precision in $\log N_{\tilde{\kappa}}$ into a required EW precision when planning observations. The same expressions apply to the recommended estimator $N_{\min}$, with $\tilde{\kappa}$ replaced by $\tilde{\kappa}_{\mathrm{box}}$ in the definition of $\gamma$. The corresponding slopes are far shallower in the saturated regime – $\gamma_{\mathrm{box}} \simeq -2.0, -2.6, -3.2$, and $-4.5$ at the same reference ratios – reflecting the logarithmic growth of $\tilde{\kappa}_{\mathrm{box}}$: The precision of $N_{\min}$ degrades only mildly as $R \to 1$, whereas that of $N_{\tilde{\kappa}}$ collapses.

The linearization fails for strongly saturated systems, where $|\gamma|\,\sigma_R/R \gtrsim 1$ ($\sigma_R$ being the propagated uncertainty of the doublet ratio), and the confidence intervals become strongly asymmetric; for the observed sample, we therefore propagate the uncertainties by Monte Carlo sampling through the exact mapping (Section 3.3). A robust one-sided limit also follows directly from the monotonicity of $\tilde{\kappa}(R)$: The upper bound on the doublet ratio provides a lower bound on the column density,

$$N_{\tilde{\kappa}} > N^{\mathrm{thin}}\,\tilde{\kappa}(R + z\,\sigma_R) \tag{A3}$$

at the confidence level corresponding to the Gaussian quantile $z$ (the subdominant uncertainty of the $N^{\mathrm{thin}}$ prefactor is neglected in this bound). The same one-sided limit applies with $\tilde{\kappa} \to \tilde{\kappa}_{\mathrm{box}}$, in which case it is fully distribution-free: a lower limit on the column density valid for an arbitrary velocity structure at the stated confidence.

This paper has been typeset from a TEX/LATEX file prepared by the author.